\documentclass[mathleft,a4paper]{an}

\Pagespan{1}{}
\Yearpublication{2012}
\Yearsubmission{2012}
\Month{12}
\Volume{333}
\Issue{10}
\DOI{10.1002/asna.2012xxxxx}

\usepackage{graphicx}
\usepackage{times}
\usepackage[scaled=0.92]{helvet}
\usepackage[usenames, dvipsnames]{color}
\usepackage{url}
\usepackage[modulo,switch]{lineno}
\usepackage{amssymb}
\usepackage{nicefrac}
\definecolor{myRed}{rgb}{0.84,0.08,0.52}

\linenumbers

\newcommand\tsp{$\;\!$}
\newcommand\phn{\phantom{0}}

\newcommand{\EDT}[1]{\textcolor{myRed}{#1}}

\usepackage[authoryear]{natbib}
\bibpunct{(}{)}{;}{a}{}{,}

 \usepackage{hyperref}
 \hypersetup{
     final=true,
     pageanchor=true,
     hyperfigures=false,
     colorlinks=true,
     breaklinks=true,
     linkcolor=blue,
     citecolor=blue,
     urlcolor=blue,
     pdfpagemode=UseNone,
     pdfpagelayout=OneColumn,
     pdfview=FitBH,
     pdftitle={The GREGOR Fabry-P\'erot Interferometer},
     pdfauthor={Puschmann et al.},
     pdfsubject={AN Special Issue 333}}

\begin{document}


\title{The GREGOR Fabry-P\'erot Interferometer}
\author{K.G.~Puschmann\inst{1}\fnmsep\thanks{Corresponding author:
    \email{kgp@aip.de}\newline},
    C.~Denker\inst{1},
    F.~Kneer\inst{2},
    N.~Al Erdogan\inst{3},
    H.~Balthasar\inst{1},
    S.M.~Bauer\inst{1},
    C.~Beck\inst{4},
    N.~Bello Gonz\'alez\inst{5},
    M.~Collados\inst{4},
    T.~Hahn\inst{1},
    J.~Hirzberger\inst{6},
    A.~Hofmann\inst{1},
    R.E.~Louis\inst{1},
    H.~Nicklas\inst{2},
    O.~Okunev\inst{7},
    V.~Mart\'inez Pillet\inst{4},
    E.~Popow\inst{1},
    T.~Seelemann\inst{8},
    R.~Volkmer\inst{5},
    A.D.~Wittmann\inst{2}, \and
    M.~Woche\inst{1}}
\titlerunning{GREGOR Fabry-P\'erot Interferometer}
\authorrunning{K.G.~Puschmann et al.}
\institute{Leibniz-Institut f\"ur Astrophysik,
    An der Sternwarte 16,
    14482 Potsdam,
    Germany
\and
    Institut f\"ur Astrophysik,
    Georg-August-Universit{\"a}t G{\"o}ttingen,
    Friedrich-Hund-Platz 1,
    37077 G\"ottingen,
    Germany
\and
    Department of Astronomy and Space Sciences, University Istanbul,
    34134  Vezneciler/Istanbul,
    Turkey
\and
    Instituto de Astrof\'{\i}sica de Canarias,
    C/ V\'{\i}a L\'actea s/n,
    38205 La Laguna,
    Tenerife,
    Spain
\and
    Kiepenheuer-Institut f{\"u}r Sonnenphysik,
    Sch{\"o}neckstra{\ss}e 6,
    79104 Freiburg,
    Germany
\and
    Max-Planck-Institut f{\"u}r Sonnensystemforschung,
    Max-Planck-Stra{\ss}e 2,
    37191 Katlenburg-Lindau, Germany
\and
    Central Astronomical Observatory, Russian Academy of Sciences,
    Pulkovskoye Chaussee 65/1,
    196140 St. Petersburg,
    Russia
\and 
    LaVision,
    Anna-Vandenhoeck-Ring 19,
    37081 G\"ottingen,
    Germany}

\received{\today}
\accepted{later}
\publonline{later}

    
\abstract{The GREGOR Fabry-P\'erot Interferometer (GFPI) is one of three
first-light instruments of the German 1.5-meter GREGOR solar telescope at the
Observatorio del Teide, Tenerife, Spain. The GFPI uses two tunable etalons in
collimated mounting. Thanks to its large-format, high-cadence CCD detectors with
sophisticated computer hard- and software it is capable of scanning spectral
lines with a cadence that is sufficient to capture the dynamic evolution of the
solar atmosphere. The field-of-view (FOV) of $50\arcsec \times 38\arcsec$ is well suited for quiet Sun and sunspot observations. 
However, in the vector spectropolarimetric mode the FOV reduces to $25\arcsec \times 38\arcsec$. 
The spectral coverage in the spectroscopic mode extends from 530--860~nm with a theoretical spectral 
resolution of ${\cal R}\approx 250,000$, whereas in the vector spectropolarimetric mode the 
wavelength range is at present limited to 580--660~nm. The combination of fast narrow-band imaging 
and post-factum image restoration has the potential for discovery science concerning the dynamic Sun 
and its magnetic field at spatial scales down to $\sim 50$~km on the solar surface.}
\maketitle

\section{Introduction}

The recently inaugurated 1.5-meter GREGOR solar telescope \citep{Volkmer2010, Schmidtetal2012a} 
is a new solar facility with the potential for cutting-edge
science at fundamental scales, i.e., the pressure scale-height, the photon
mean-free path, and the intrinsic size of elemental magnetic structures. 
The key science topics of the GREGOR telescope are \citep[see][]{Schmidtetal2012b}: (1) 
the interaction between convection and magnetic
fields in the photosphere, (2) the nature and dynamics of sunspots and pores and
their temporal evolution, (3) the solar magnetism and its role in solar
variability, and (4) the enigmatic heating mechanism of the chromosphere.
The GREGOR telescope is equipped with powerful post-focus instrumentation for different wavelength
regimes and science targets. The GRating Infrared Spectrograph \citep[GRIS,
][]{Colladosetal2008,Colladosetal2012} and the GREGOR Fabry-P\'erot
Interferometer (GFPI) focus on spectropolarimetry in the near infrared and
visible spectral ranges, respectively. Both instruments are accompanied by the
Broad-Band Imager \citep[BBI,][]{VDL2012} for high-cadence, high-resolution
imaging. 

\begin{figure*}
\centering
\includegraphics[width=\textwidth]{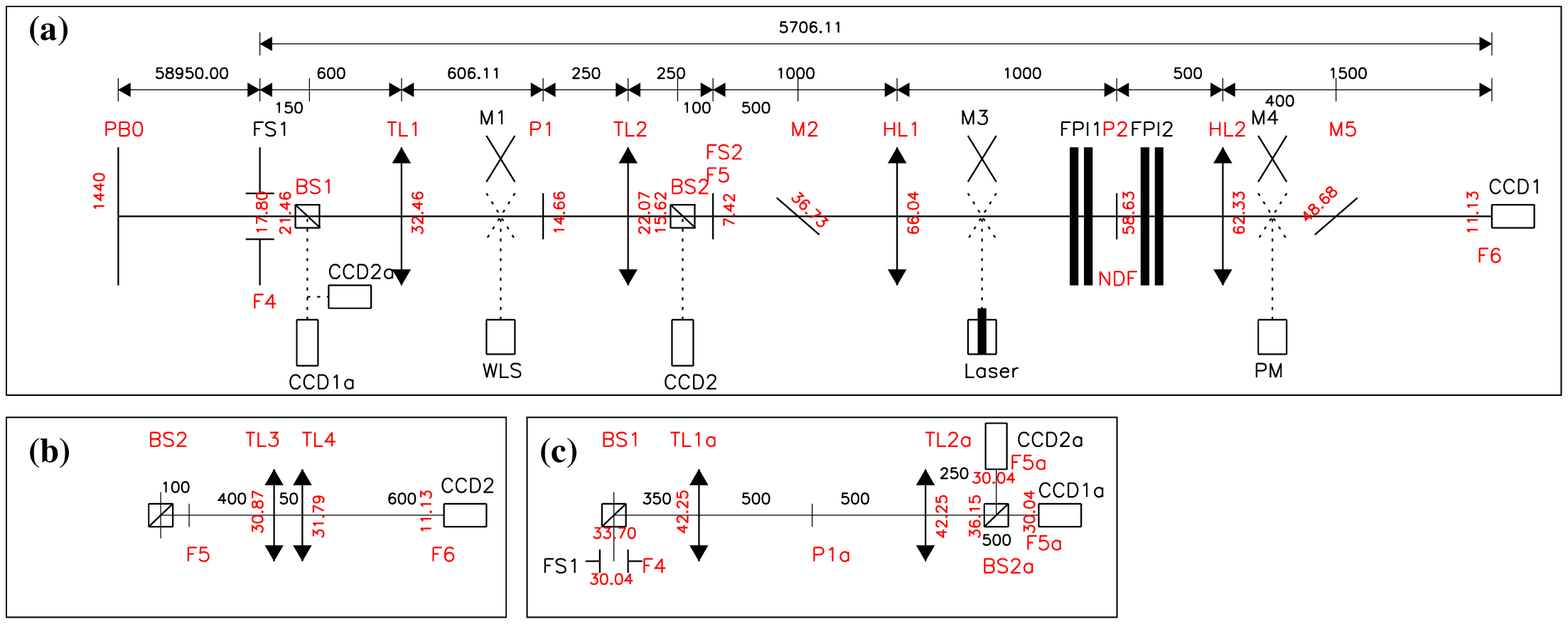}
\caption{Schematic design of the main channels of the GFPI: narrow-band channel \textsf{NBC} (a),
    broad-band channel \textsf{BBC (b)}, and blue imaging channel \textsf{BIC} (c). 
    The optical elements of the \textsf{BIC} are labeled with an extra `\textsf{a}' 
    (see text and Tab.~\ref{Tab1} for details). The auxiliary channels, i.e., a 
     laser-photomultiplier channel (\textsf{Laser}, \textsf{PM}) and a white-light channel 
     (\textsf{WLS}) are only depicted in an illustrative way. All the distances are shown in millimeters.}
\label{Fig01}
\end{figure*}

The roots of the GFPI can be traced back to 1986, when the
Universit\"ats-Sternwarte G\"ottingen started to develop an imaging spectrometer
for the German Vacuum Tower Telescope (VTT). This instrument used a universal
birefringent filter (UBF) as an order-sorting filter for a narrow-band FPI
mounted in a collimated light beam \citep{Bendlinetal1992}. The UBF was later
replaced by a second etalon \citep{Volkmeretal1995} and the spectrometer was
equipped with a Stokes-$V$ polarimeter \citep{Koschinskyetal2001}.

First ideas concerning an imaging spectrometer at the GREGOR telescope have been
published by \citet{KneerHirzberger2001} and \citet{Kneeretal2003}. A renewal of
the G\"ottingen FPI during the first half of 2005 was, however, the first
fundamental step towards the development of the GFPI \citep{Puschmannetal2006}.
This included the integration of narrow-band etalons with diameters of 70~mm and
large-format, high-cadence CCD detectors, both accompanied by sophisticated
computer hard- and software. From 2006  to 2007, the optical design of the GFPI
was finalized and all relevant optical and opto-mechanical components were
bought or manufactured \citep{Puschmannetal2007}. Subsequently, an upgrade to
full-Stokes spectropolarimetry followed \citep{BelloGonzalezKneer2008,
Balthasaretal2009}.

In 2009, the Leibniz-Institut f\"ur Astrophysik Potsdam took over the scientific
responsibility for the GFPI, and the instrument was finally installed at the
GREGOR solar telescope \citep{Denkerteal2010b}. The integration of three
computer-controlled translation stages (two filter sliders and one mirror stage)
and the preparation of the software for TCP/IP communication with external
devices according to the Device Communication Protocol \citep[DCP,
][]{Halbgewachs2012} followed during the commissioning phase in 2011
\citep{Puschmannetal2012a}.

The current state of the instrument, including advanced and automated
calibration and observing procedures and a new blue imaging channel
(380--530~nm), is described in \citet{Puschmannetal2012b}.  Beside an inspection
of first observational results obtained with the GFPI at the GREGOR telescope, the latter
publication also includes a design concerning optics, cameras, and etalons for
the future integration of the BLue Imaging Solar Spectrometer (BLISS), a second
Fabry-P\'erot for the wavelength range 380--530~nm that has been suggested by
\citet{Denker2010a}. Thus, \citet{Puschmannetal2012a, Puschmannetal2012b} and the
detailed description presented here give a complete overview of the present state of the
instrument at the end of the 2012 science verification campaigns and its
foreseen use at the GREGOR solar telescope.

Starting in 2002, many colleagues have contributed to the GFPI, which has been
developed at the Institut f\"ur Astrophysik G\"ottingen and at the
Leibniz-Institut f\"ur Astrophysik Potsdam. In the spirit of this special issue
of \textit{Astronomische Nachrichten/AN}, the extensive list of co-authors
honors all contributions to this instrument over the years. Details about the
individual contributions can be glimpsed from the articles in the above
chronology.


\begin{figure*}
\centerline{\includegraphics[width=\textwidth]{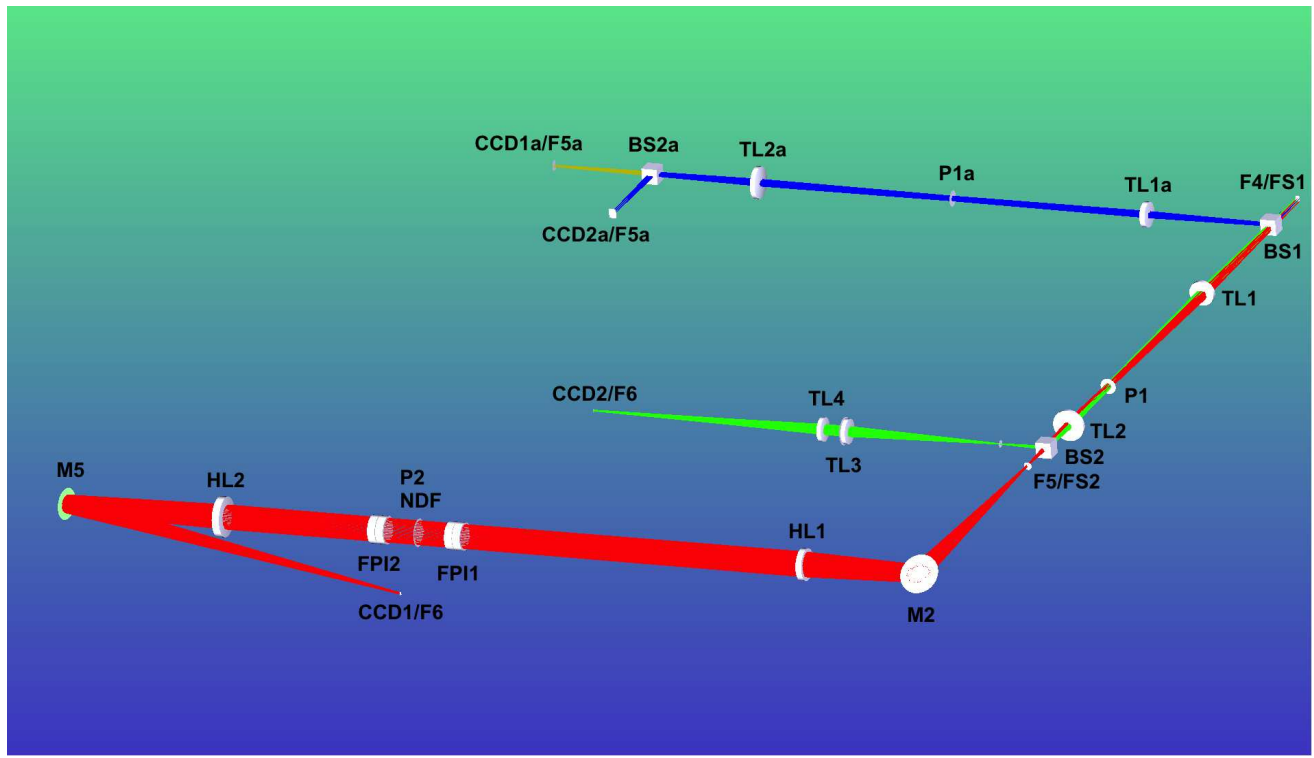}}
\caption{\textsf{NBC} (red), \textsf{BBC} (green), and \textsf{BIC} (blue) in a
    ZEMAX multi-configuration file in shaded modeling for the respective
    central wavelength and maximal field dimension of each beam. 
    The optical elements of the \textsf{BIC} are labeled with an extra `\textsf{a}' 
    (see text and Tab.~\ref{Tab1} for details).}
\label{Fig02}
\end{figure*}

\section{Optical design}

\begin{table}
\caption{Properties of the main optical components of the GFPI.}
\label{Tab1}

\parbox{50mm}{
\footnotesize
\begin{tabular}{lcc}
\hline\hline
Optical\rule{0mm}{3mm}       & Focal length & Diameter \\
element\rule[-1mm]{0mm}{3mm} & [mm]         & [mm] \\
\hline
TL1\rule{0mm}{3mm}           & \phn 600     & \phn 63 \\
TL2                          & \phn 250     & \phn 40 \\
HL1                          &     1000     & \phn 80 \\
HL2                          &     1500     &     100 \\
TL3                          & \phn 400     & \phn 63 \\
TL4                          & \phn 600     & \phn 63 \\
TL1a                         & \phn 500     & \phn 63 \\
TL2a\rule[-1mm]{0mm}{3mm}    & \phn 500     & \phn 80 \\
\hline
\end{tabular}}
\parbox{20mm}{
\footnotesize
\vspace{-10mm}
\begin{tabular}{lc}
\hline\hline
Optical\rule{0mm}{3mm}        & Dimension \\
element\rule[-1mm]{0mm}{3mm}  & [mm $\times$ mm]\\
\hline
BS1\rule{0mm}{3mm}            & $40 \times 40$ \\
BS2                           & $40 \times 40$ \\
BS2a                          & $40 \times 40$ \\
M2                            & $60 \times 85$ \\
M5\rule[-1mm]{0mm}{3mm}       & $60 \times 85$ \\
\hline
\end{tabular}}
\end{table}

The GFPI is fed with light from the telescope by a dichroic pentaprism that is
located behind the adaptive optics just before the science focus \textsf{F4}.
The pentaprism transmits the infrared wavelength range towards GRIS and reflects
the visible wavelength range towards GFPI. The currently available pentaprism
splits the light at 660~nm and thus permits GFPI observations at wavelengths
shorter than H$\alpha$. In the near future, a second pentaprism will enable
observations of H$\alpha$ and the Ca\,\textsc{ii} infrared lines around 850~nm,
although the suitability of the GFPI cameras for the latter has still to be
investigated because of the detector's low quantum efficiency ($<15$\%) at
infrared wavelengths (see Sect.~\ref{cameras}).

Figure~\ref{Fig01} shows a schematic design of the current setup of the main
components of the GFPI, i.e., the narrow-band channel (\textsf{NBC}, panel a),
the broad-band channel (\textsf{BBC}, panel b), and the blue imaging channel
(\textsf{BIC}, panel c). This design is based on calculations following
geometrical optics, a strategy that was also the base for the original design of
the GFPI presented in \citet[][their Fig.~1]{Puschmannetal2007}. The diameter of
the light beam at the different foci, pupils, and on the relevant optical
surfaces is indicated together with an illustration of the auxiliary beams of
the instrument. The focal lengths and diameters of individual lenses and the
dimensions of other optical elements are summarized in Tab.~\ref{Tab1}. As in 2007, 
the calculations based on geometrical optics were confirmed by a ZEMAX ray-tracing as
shown in Fig.~\ref{Fig02}.

\begin{figure}
\centering
\includegraphics[width=\columnwidth]{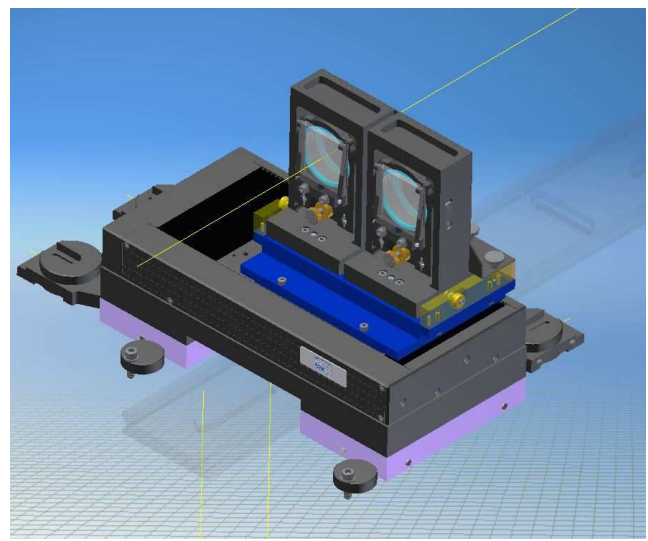}
\caption{CAD of the filter sliders mounted in NBC and BBC.}
\label{Fig03}
\end{figure}

The entire optical setup of the GFPI, including details about the auxiliary
channels, i.e., the laser-photomultiplier channel and the white-light channel
for etalon adjustment and spectral calibration purposes, as well as the
distribution of all optical elements on the optical tables, can be found in a
drawing up to scale in \citet[][their Fig.~1]{Puschmannetal2012b}.

\begin{figure*}
\begin{center}
\includegraphics[width=\columnwidth]{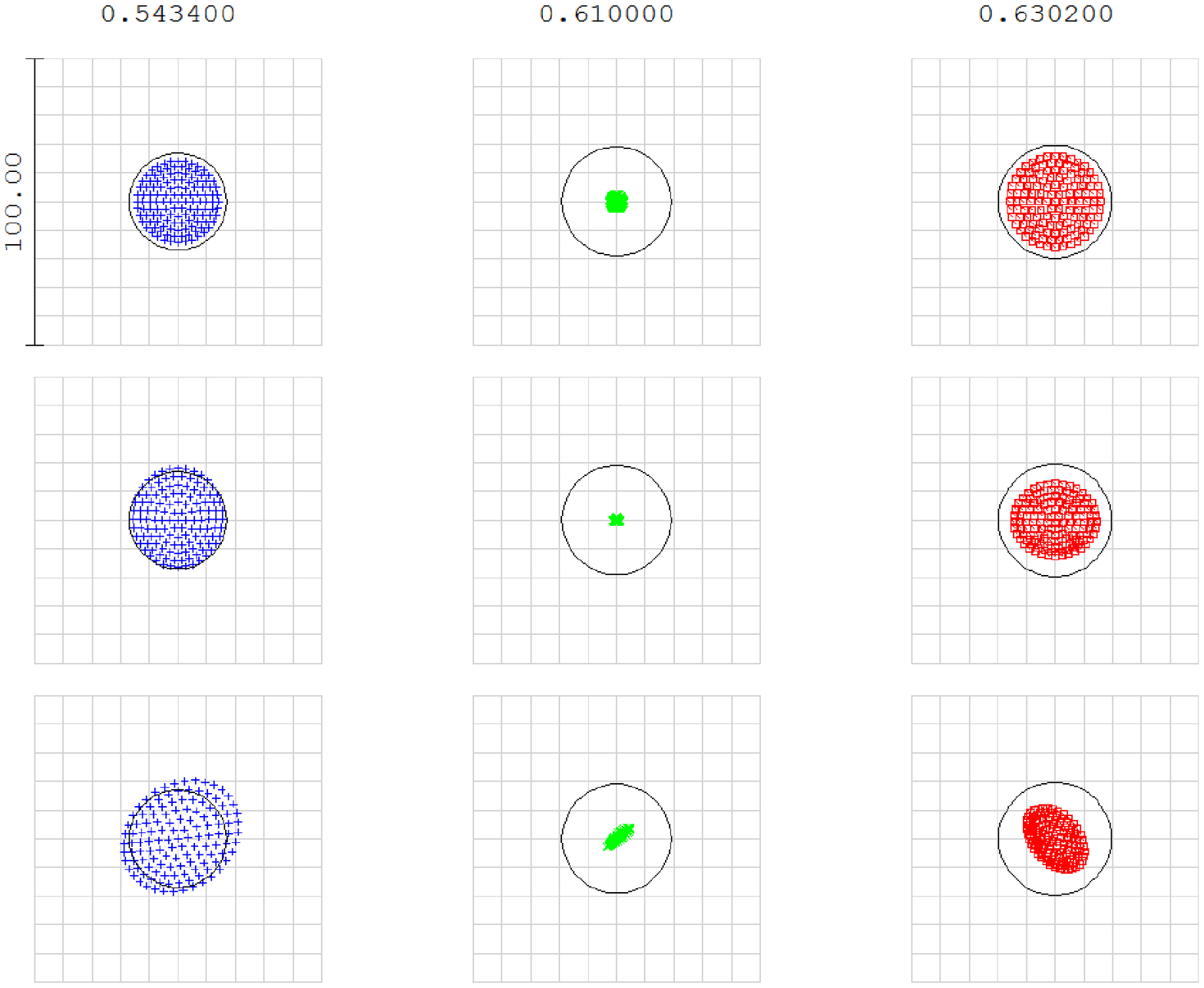}
\includegraphics[width=\columnwidth]{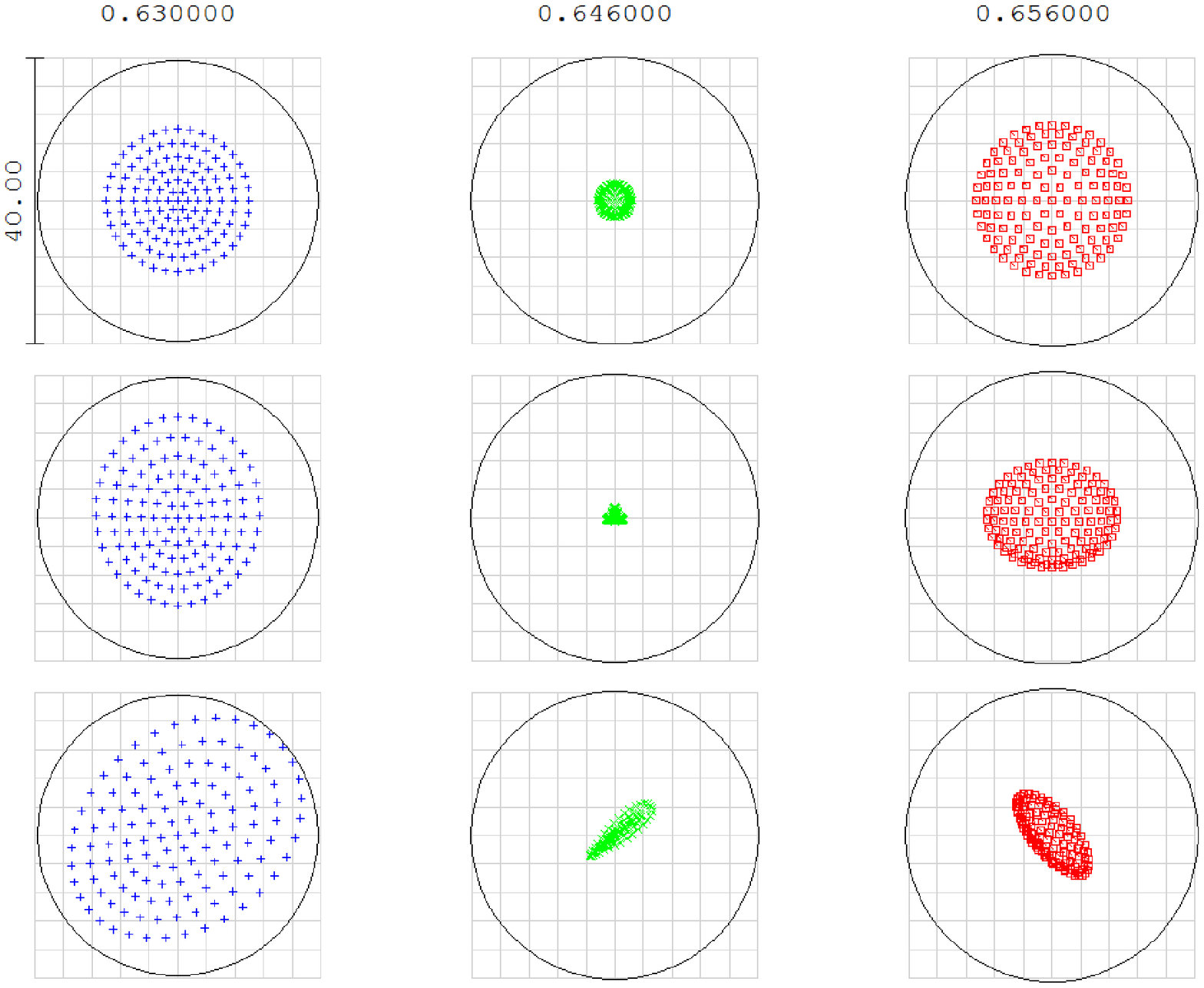}
\end{center}
\caption{Matrix spot diagrams of the \textsf{NBC} for different wavelengths and
    fields. In each panel the top row represents the optical axis, whereas the
    the middle and bottom row represents a field point at the middle right and 
    in the diagonal corner on the detector, respectively. Left panel: 
    wavelength range 543.4--630.2~nm with 610~nm in focus. Right panel: wavelength range 630--656~nm with 646~nm in
    focus. All spots inside the Airy-disk (circles in the boxes) are diffraction
    limited. The box sizes of 100~$\mu$m and 40~$\mu$m correspond to
    0\farcs56 and 0\farcs22, respectively.} 
    \label{FIG05}
\end{figure*}

In the \textsf{NBC} of the GFPI, the science focus \textsf{F4} is re-imaged
twice by the achromatic lenses \textsf{TL1}, \textsf{TL2}, \textsf{HL1}, and
\textsf{HL2}. This creates two foci (\textsf{F5} and \textsf{F6}) and two pupil
images (\textsf{P1} and \textsf{P2}). The re-imaging is used to achieve a proper
size of the pupil image \textsf{P2} that is limited by the free aperture of the
two etalons (\textsf{FPI1} and \textsf{FPI2}), which are mounted close to the
pupil image in a collimated light beam. A neutral density filter 
(\textsf{NDF}) of high optical quality between \textsf{FPI1} and \textsf{FPI2} 
and with a transmission of 63\% reduces the inter-etalon reflexes. 
The beam in \textsf{NBC} is folded twice by the two deflection 
mirrors \textsf{M2} and \textsf{M5}. Two field stops \textsf{FS1} and 
\textsf{FS2} prevent an over-illumination of the detectors. Furthermore, 
\textsf{FS2} can be adjusted to avoid an overlap of the two images created 
by the removable dual-beam full-Stokes polarimeter.  
A beam splitting cube \textsf{BS2} in front of \textsf{F5} deflects 
5\% of the light into the \textsf{BBC}, where \textsf{F5} is re-imaged 
by the achromatic lenses \textsf{TL3} and \textsf{TL4}. 

In the \textsf{NBC} and \textsf{BBC}, two computer-controlled filter sliders
(see Fig.~\ref{Fig03}) allow one to sequentially observe two different
wavelength bands. The filters restrict the bandpass for \textsf{BBC} and
\textsf{NBC} to a full-width-at-half-maximum (FWHM) of 10~nm and 0.3--0.8~nm,
respectively. Narrow-band filters with a FWHM in this range provide
 a good compromise between parasitic light, transmission, and signal-to-noise. 
The pre-filters of each channel can be tilted to optimize the wavelength of 
the transmission maximum. The sliders are limited to two interference filters 
because of the free space available in the \textsf{NBC}. The near-focus position 
of the filter sliders additionally required a high precision in the 
positioning of the pre-filters down to a few micrometers, which would 
have not been achievable with standard filter wheels. First science 
verification campaigns at GREGOR revealed the need for very long exposure 
times (up to 100~ms) in the \textsf{NBC} at full resolution of the cameras, 
when using filters with a transmission $T \approx 40$\%, e.g., for spectral scans 
of the Fe\,\textsc{i} $\lambda 543.4$~nm and Fe\,\textsc{i} $\lambda 557.6$~nm lines. 
A $2 \times 2$-pixel binning of the
CCDs reduces significantly the exposure times and increases the frame rate from
7 to 16 frames~s$^{-1}$. Filters with $T \approx 80$\%, e.g., for spectral scans
of the Fe\,\textsc{i} $\lambda 617.3$~nm and Fe\,\textsc{i} $\lambda 630.2$~nm
lines, yield reasonable frame rates and exposure times even without binning
\citep[see][]{Puschmannetal2012b}.

In both \textsf{NBC} and \textsf{BBC}, Sensicam QE CCD detectors (\textsf{CCD1}
and \textsf{CCD2}) with an identical image scale at \textsf{F6} are used for
simultaneous data acquisition. Co-temporal recorded broad- and
narrow-band data are fundamental for a post-facto image restoration, which
improves the spatial resolution of the data far beyond the real-time correction
provided by the GREGOR Adaptive Optics System \citep[GAOS,
][]{Berkefeldetal2012}. A data pipeline is nearly finished \citep[see
][]{Puschmannetal2012b}, which includes both speckle and blind deconvolution
reconstruction methods. A comparison of different methods, their application,
and their results is given by \citet{PuschmannBeck2011}.

A pentaprism \textsf{BS1} deflects all wavelengths below 530~nm into the
\textsf{BIC}, where \textsf{F4} is re-imaged by two achromatic lenses
(\textsf{TL1a} and \textsf{TL2a}). A 50/50 beam splitting cube finally directs
the light onto two pco.4000 cameras that are in stock at the observatory.
Different available pre-filters, e.g., at Ca\,\textsc{ii}\,H $\lambda
396.8$~nm, the Fraunhofer G-band $\lambda 430.7$~nm, and a blue continuum window
$\lambda 450.6$~nm with a $\mathrm{FWHM} = 1$~nm and $T = 60$\%, can be placed
in front of these two cameras for complementary high-cadence imaging.

All achromats used in the GFPI have been purchased off-the-shelf. Thus, the
instrument is not super-achromatic. Observations in two different wavelength
bands are restricted to a maximum wavelength separation of about 100~nm.
Figure~\ref{FIG05} shows matrix spot diagrams of the \textsf{NBC} for the case
of multi-wavelength observations in the wavelength bands 543.4--630.2~nm (left
panel) and 630--656~nm (right panel) with the cameras focused at 610 and 646~nm,
respectively. Both panels exhibit the typical behavior of achromatic lens
systems, when observing at wavelengths either to the red or blue of the
wavelength in focus. At the edge wavelengths, the dots show a larger spread on-axis 
(top row of Fig.~\ref{FIG05}), which is accompanied by a directional component 
off-axis (bottom row of Fig.~\ref{FIG05}). For a large wavelength separation 
from the central wavelength in focus, the spread of the dots exceeds the Airy disk 
and the performance is no longer diffraction-limited (e.g., lower leftmost panel). 
Modern achromatic lenses such as those of Quioptiq used in the GFPI are matched for 
the H$\alpha$ $\lambda 656.3$~nm and H$\beta$ $\lambda 486.1$~nm lines with a remaining 
variation of the focus position at 546~nm typically below 0.1\%. As a consequence, 
we find a chromatic focus shift when observing away from the wavelength in focus. 
The different imaging quality between the left and the right panel is caused by the 
smaller spectral range used in the right panel and the stronger non-linearity of the 
chromatic focal shift in the spectral range for the left panel.

\section{Cameras and control software}


\label{cameras}

\begin{figure}
\begin{center}
\includegraphics[width=\columnwidth]{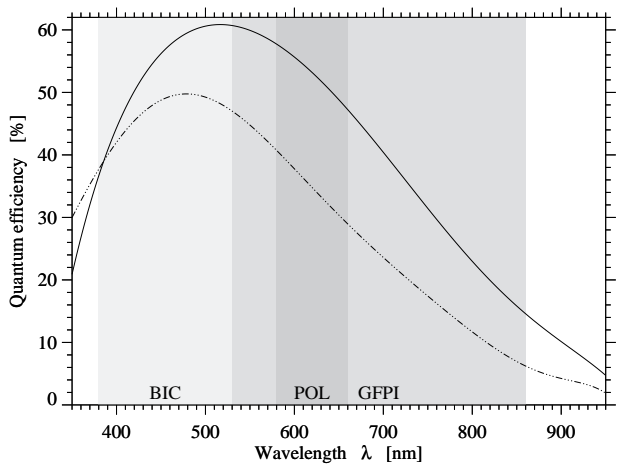}
\end{center}
\caption{Quantum efficiency of the Sensicam~QE CCD (\textit{solid}) and pco.4000
     (\textit{dash-dotted}) cameras. The different grey levels denote the respective 
     wavelength ranges for the \textsf{GFPI} in spectroscopic (grey) and 
     vector-polarimetric mode (dark grey), and for the blue imaging channel \textsf{BIC} (light grey).}
\label{FIG04}
\end{figure} 

\textsf{NBC} and \textsf{BBC} are equipped with Sensicam QE CCD cameras with an
analog-digital conversion with 12-bit resolution. Their Sony ICX285AL detectors
have a full-well capacity of 18,000~e$^{-}$ and a read-out noise of 4.5~e$^{-}$.
Their quantum efficiency (QE) is depicted in Fig.~\ref{FIG04} with a maximum of 
$\approx 60$\% at 550~nm. The detectors have $1376 \times
1040$ pixels with a size of 6.45~$\mu$m $\times$ 6.45~$\mu$m. The spatial
sampling per pixel is 0\farcs0361 \citep[see][their
Tab.~3]{Puschmannetal2012b}. Consequently, a field-of-view (FOV) of $50\arcsec
\times 38\arcsec$ is available in the spectroscopic mode of the instrument.

The cameras have a maximum frame rate at full resolution of 10~Hz. As mentioned
before, long exposure times can further reduce the read-out frequency. This is
especially problematic when using the instrument in full-Stokes mode. There, the
duration of spectral line scans can increase to a few minutes when repeated
images in the four different polarimetric states at up to 80 different
wavelength positions are taken to improve the signal-to-noise ratio. Therefore,
it is presently recommended to use these cameras with a $2\times2$-binning for
such data, which still yields a spatial resolution of up to 0\farcs14. By this,
the intensity increases by a factor of four, resulting in shorter exposure times
and much higher frame rates of up to 20~Hz. Polarimetry with the GFPI would
certainly benefit by a future replacement of these cameras by modern
sCMOS cameras with frame rates of up to 40~Hz without any binning
\citep[see][]{Puschmannetal2012b}.

For imaging in the \textsf{BIC}, two pco.4000 cameras are available at the
observatory. Their detectors have a full-well capacity of 60,000~e$^{-}$ and a
read-out noise of 11~e$^{-}$. The frame rate of these cameras is limited to 2.3~Hz at 
full chip size using only one analog-digital converter.
The QE is shown in Fig.~\ref{FIG04}, with $\approx 32$\% ($\approx 45$\%) at a wavelength of 380 (530)~nm. 
With a pixel size of 9~$\mu$m $\times$ 9~$\mu$m, their $4008
\times 2672$ pixels yield a total chip size of 36~mm $\times$ 24~mm. The image
scale at both cameras is 0\farcs0315\ pixel$^{-1}$. A field stop in \textsf{F4} avoids a 
vignetting of the blue beam. The illuminated part of the sensor thus corresponds to only 
$2000 \times 2672$ pixels, yielding a FOV of $63\arcsec \times 84\arcsec$. 
With the reduced read-out area on the chip, the cameras permit the recording of more than 30 images in 
less than 15~s, which is sufficient to allow for post-facto image reconstruction. 
However, \textsf{BIC} is just an intermediate solution and will be replaced by the 
Blue Imaging Solar Spectrometer \citep[BLISS,][]{Puschmannetal2012b} 
in the near future. A summary of the properties of the GFPI image acquisition 
systems can be found in Tab.~\ref{Tab2}.

\begin{table}[t]
\caption{Properties of the GFPI image acquisition system.}
\label{Tab2}
\begin{tabular}{lrr}
\hline\hline
 & NBC and BBC  & BIC\rule[-1.5mm]{0mm}{5mm}   \\
\hline
Camera\rule{0mm}{3.5mm}       & 2$\times$ Imager QE  & 2$\times$ pco.4000\\
Detector type                 & CCD                  & CCD            \\
Pixels                        & $1376 \times 1040$   & $2000 \times 2672$ \\
Pixel size [$\mu$m$^2$]       & $6.45 \times 6.45$   & $9.0 \times 9.0$ \\
Read out noise [e$^{-}$]      & 4.5                  &    11            \\
Full well capacity [e$^{-}$]  & 18,000               &  60,000          \\
Spectral resp. [nm]           & 320--900             &  320--900       \\
Quantum efficiency            & 60\% @ 550~nm        &  32\% @ 380~nm   \\
                              & 45\% @ 530~nm        &  54\% @ 530~nm   \\
Digitization [bit]            & 12                   &  14              \\
FOV                           & $50\arcsec\times 38\arcsec$ &
    $63\arcsec\times 84\arcsec$ \\
Image scale                   & 0\farcs0361 pixel$^{-1}$ & 0\farcs0315
   pixel$^{-1}$ \\
Frame rate [Hz] \rule[-1.5mm]{0mm}{3mm}  & 10        &  2.3\\
\hline
\end{tabular}
\end{table}

\begin{table}[t]
\caption{Etalon- and spectral properties of the GFPI.}
\label{Tab3}
\begin{tabular}{lcc}
\hline\hline
                           &  Etalon 1\rule[-1.5mm]{0mm}{5mm}            &
    Etalon 2            \\
\hline
Manufacturer\rule{0mm}{3.5mm} &  ICOS  & ICOS  \\
Diameter [mm]              &  70                  & 70                  \\
Finesse                    &  46                  & 46                  \\
Reflectivity [\%]          &  95                  & 95                  \\
Etalon spacing [mm]        &  1.1                 & 1.4                 \\
FWHM [pm] @~617~nm        &  3.43                & 2.73                 \\
FWHM combined [pm] @~617~nm  &  \multicolumn{2}{c}{1.95}                 \\
Spectral resolution        &  \multicolumn{2}{c}{250,000}\\
Parasitic-light fraction [\%] &  \multicolumn{2}{c}{$\approx 1.0$}\\
Controllers                &  \multicolumn{2}{c}{CS-100 (ICOS)}       \\
Spectral scanning          &  \multicolumn{2}{c}{RS-232}              \\
Coating\rule[-1.5mm]{0mm}{3mm} &  \multicolumn{2}{c}{530--860~nm}        \\
\hline
\end{tabular}\\
\footnotesize \hspace*{5mm}
\parbox{70mm} {\vspace*{-2mm} 
\begin{itemize}
\item[Note:] The parasitic-light fraction is given 
for a pre-filter with $\mathrm{FWHM}=0.3$~nm and is valid for the entire wavelength range. 
\end{itemize}}
\end{table}

The communication between internal (cameras, etalons, polarimeter, and filter and
mirror sliders) and peripheral devices (telescope, AO system, AO filter wheel,
GPU, GRIS, etc.) is controlled by the software package DaVis from LaVision in
G\"ottingen. Its first adaptation to the spectrometer, including a description
of the resulting Graphical User Interface (GUI) and required scan tables, is
described in detail in \citet{Puschmannetal2006}. The modification of the
software for TCP/IP communication with external devices using DCP and the
subsequent implementation of automated observing and calibration procedures is
depicted in \citep{Puschmannetal2012a, Puschmannetal2012b}. \citet[][their
Fig.~2]{Puschmannetal2012a} shows a flow chart of the communications of the GFPI
control computer (DaVis) with all internal and peripheral devices.

\section{Dual etalon system}

The central optical element of the GFPI is a dual-etalon system, which is
mounted in the collimated beam close to a pupil image with a 58~mm diameter. The
etalon properties and theoretical spectral characteristics of the instrument are
summarized in Tab.~\ref{Tab3} and are described in detail below.

The normalized transmission profile of a single etalon is given by
\begin{eqnarray}
\frac{I}{I_0} & = & \frac{T}{1+F\sin^2\nicefrac{\delta}{2}}\label{TRANS}\\
T             & = & \left(1 - \frac{A}{1-R} \right)^2\label{TMAX}\\
F             & = & \frac{4 R}{(1-R)^2}\\
\delta        & = & \frac{4 \pi n d \cos \theta}{\lambda} + 2 \Phi\label{DELTA}
\end{eqnarray}
with the  incident intensity $I_0$, transmitted intensity $I / I_0$,
reflectivity $R$, absorption $A$, peak transmission $T$, phase difference
$\delta$, index of refraction $n$, plate spacing of the etalons $d$, angle of
incidence $\theta$, phase change on internal reflections $\Phi$, and wavelength
$\lambda$.

\begin{figure}[t]
\centering
\includegraphics[width=\columnwidth]{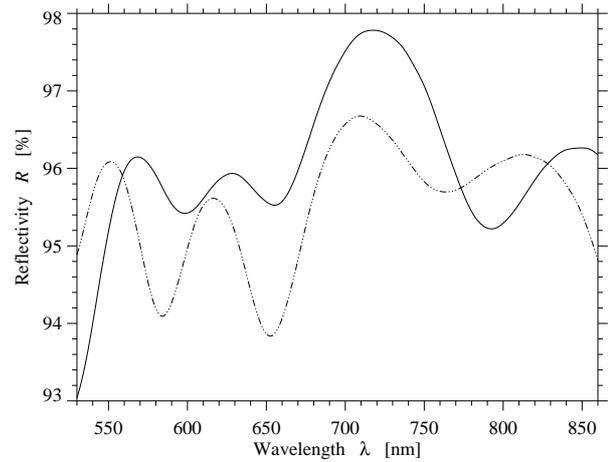}
\caption{Reflectivity $R_1$ of the broad-band \mbox{(\textit{dash-dotted}\tsp)}
    and $R_2$ of the narrow-band \mbox{(\textit{solid}\tsp)} etalon coating
curves.}
\label{FIG07}
\end{figure}

\begin{figure}[t]
\centering
\includegraphics[width=\columnwidth]{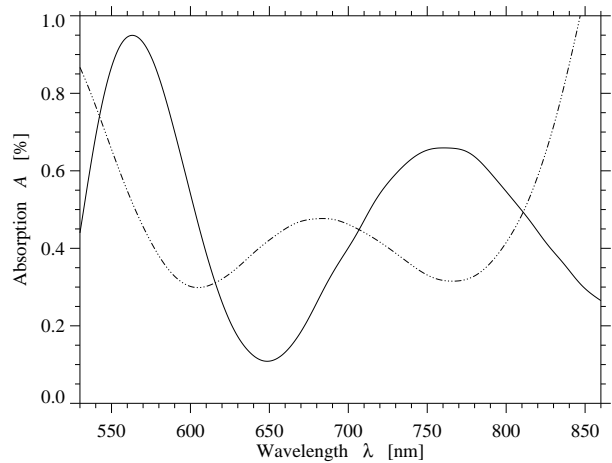}
\caption{Absorption $A_1$ of the broad-band \mbox{(\textit{dash-dotted}\tsp)}
    and $A_2$ of the narrow-band \mbox{(\textit{solid}\tsp)} etalon coating
curves.}
\label{FIG06}
\end{figure}

The plate spacings of the two narrow-band etalons are $d_1 = 1.100$~mm and 
$d_2 = 1.408$~mm because of historical reasons (see below) 
and were provided by IC Optical Systems (ICOS) with an accuracy of $\pm 1$~$\mu$m. 
Somewhat larger deviations from the nominal values were
observed, while determining the precise step ratios, which are required to tune
the  etalons in tandem. However, these small deviations are of no consequence
for the following discussion of the etalon characteristics. The manufacturer
also furnished the coating curves for both etalons. The reflectivity $R_{1,2}$
and absorption $A_{1,2}$ curves are shown in Figs.~\ref{FIG07} and \ref{FIG06},
respectively. A nominal reflectivity $R = 96$\% and absorption $A = 0.5$\% was
specified as the goal for the wavelength range 530--860~nm. In practice, flat
coating curves covering such a broad wavelength range cannot be achieved. In
general, the deviations from the nominal values are small. Only toward the blue
and infrared stronger departures from the nominal values are observed, so that
observations outside the bounds of the wavelength region are not impossible but
the performance of the etalons will decrease significantly.

\begin{figure}[t]
\centering
\includegraphics[width=\columnwidth]{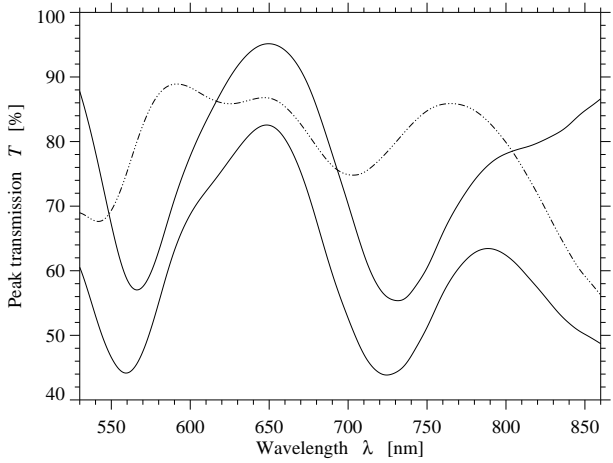}
\caption{Peak transmission $T_1$ of the broad-band
    \mbox{(\textit{dash-dotted}\tsp)} and $T_2$ of the narrow-band
    \mbox{(\textit{thin solid}\tsp)} etalon and the combined peak transmission $T$
    \mbox{(\textit{thick solid}\tsp)} of the dual-etalon system.}
\label{FIG08}
\end{figure}

The peak transmissions $T_{1,2}$ depend only on the reflectivities $R_{1,2}$
and absorptions $A_{1,2}$ of the two etalons (Eq.~\ref{TMAX}). The combined
peak transmission $T$ is simply the product of $T_1$ and $T_2$. The
corresponding transmission curves in Fig.~\ref{FIG08} demonstrate clearly that
FPIs can achieve nowadays a very high transmission, i.e., 60--90\% in case of a
single etalon and 40--80\% for a dual-etalon system. The peak transmission of
the GFPI exceeds 60\% in the wavelength range 590--690~nm. The highest
transmission of more than 80\% is achieved near the strong chromospheric
absorption line H$\alpha$ $\lambda$656.28~nm.

\begin{equation}
{\cal F} = \frac{\mathrm{FSR}}{\mathrm{FWHM}} = \frac{\pi \sqrt{R}}{1-R}
\label{FIN}
\end{equation}
The reflectivity finesse ${\cal F}$ also enters in the Airy function
(Eqs.~\ref{TRANS}--\ref{DELTA}) because of its relation to the parameter $F$.
\begin{equation}
{\cal F}^2 = \frac{\pi^2 F}{4}
\end{equation}
However, in real applications, higher order effects have to considered. Before
coating, the etalon surfaces have an optical quality of $\lambda / p = \lambda /
200$ as specified by the manufacturer (with $p$ being the plate-defect finesse). After coating $p$ reduces to 140. The 
nominal finesse ${\cal F}_\mathrm{nom}$ includes the contributions by plate defects.
\begin{equation}
{\cal F}_\mathrm{nom} = \left[ \frac{(1-R)^2}{\pi^2 R} + \frac{4}{p^2}
    \right]^{-\nicefrac{1}{2}}
\end{equation}

\begin{figure}
\begin{center}
\includegraphics[width=\columnwidth]{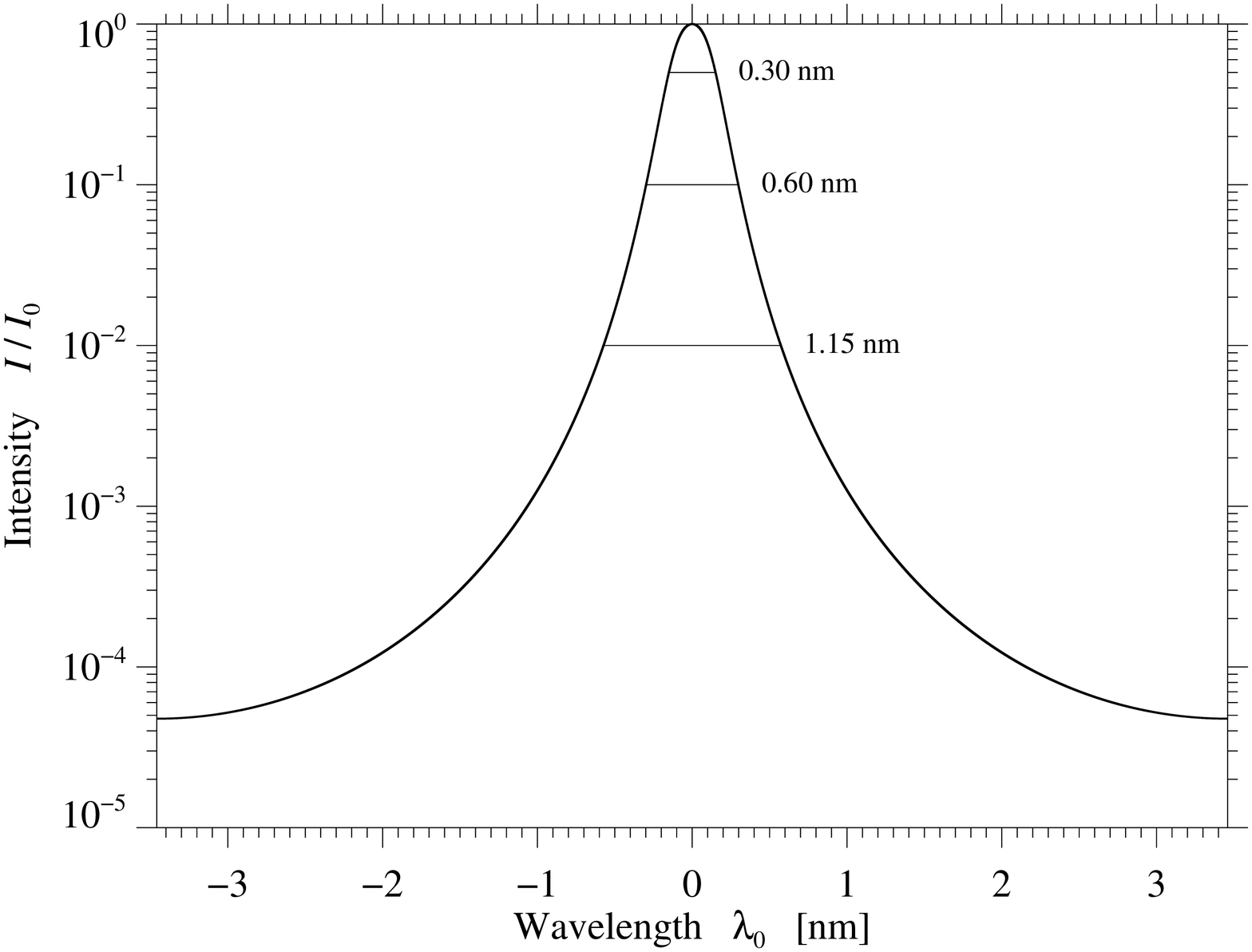}
\includegraphics[width=8cm]{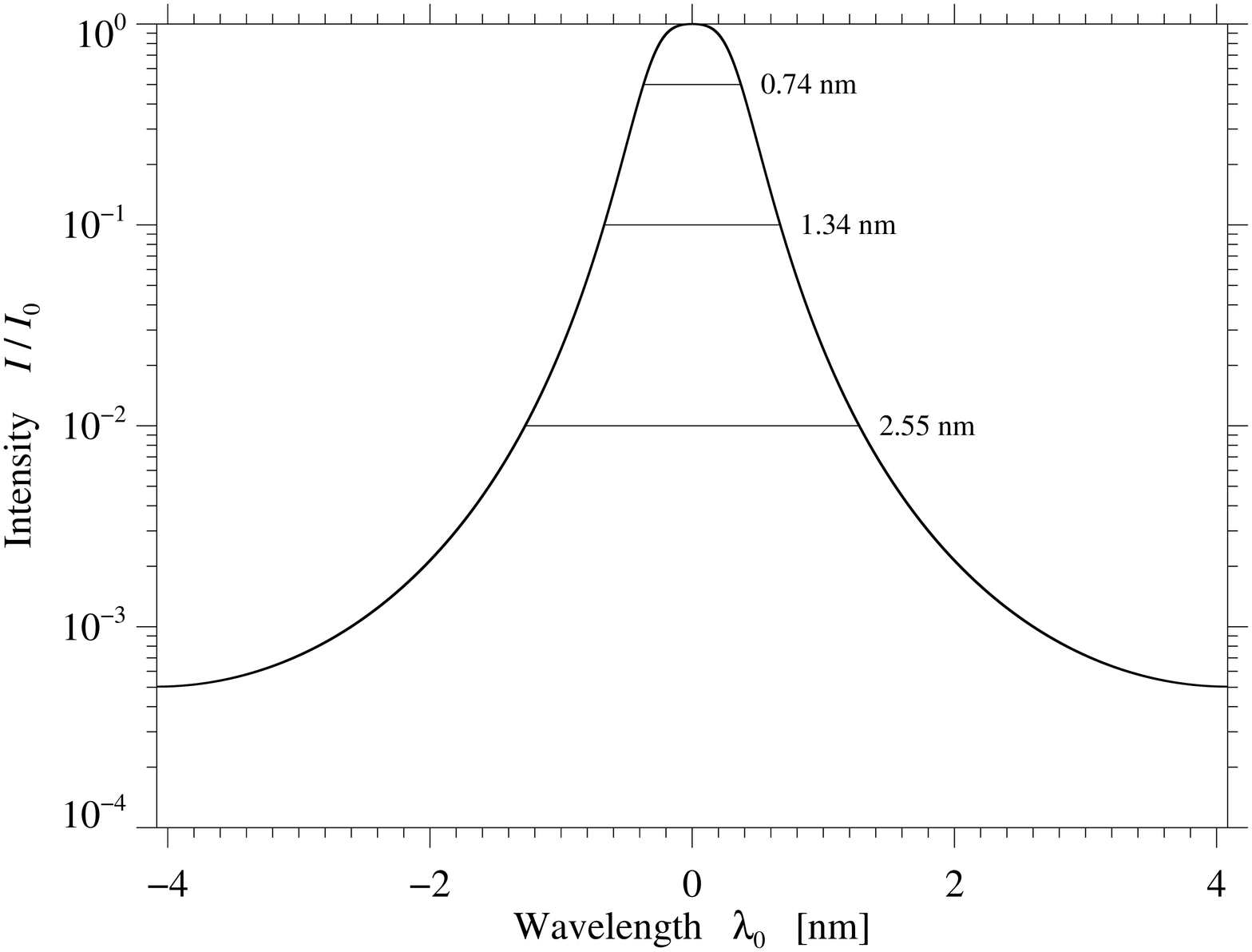}
\end{center}
\caption{Transmission profiles of two-cavity interference filters with a FWHM
    of 0.30~nm \mbox{(\textit{upper panel})} and 0.74~nm \mbox{(\textit{lower panel})}, which
    are representative for the GFPI pre-filters. The interference
    filters consist of two superposed Airy functions with a reflectivity $R =
    0.87$ and 0.80 and cavity spacing $d = 26$~$\mu$m and 22~$\mu$m, which are
    displaced by $\pm 0.10$~nm and $\pm 0.26$~nm from the central wavelength
    $\lambda_0$. Since the combined Airy functions are periodic, they were
    plotted only between the minima enclosing the maximum at the central
    wavelength $\lambda_0$. The labels in the plot refer to the width of the
    interference filters at normalized intensities $I / I_0 = 0.5$, 0.1, and
    0.01, respectively.}
\label{FIG09}
\end{figure}

According to Eqs.~\ref{TRANS} and \ref{DELTA}, the transmission curve of an
etalon is a quasi-periodic function. Even in a dual-etalon system,
interference filters have to be introduced as order-sorting filters. A
dual-cavity interference filter can be constructed using the superposition of
two etalons with fixed but almost identical plate spacings, which are slightly
displaced from the central wavelength $\lambda_0$. The superposition of two Airy
functions (Eqs.~\ref{TRANS}--\ref{DELTA}) yields a good approximation of the
custom interference filters used at the GFPI, where the details of the
manufacturing process are proprietary. The two transmission curves shown in
Fig.~\ref{FIG09} were constructed to resemble the Fe\,\textsc{i} $\lambda
543.4$~nm and Fe\,\textsc{i} $\lambda 617.3$~nm GFPI filters, which have a FWHM
of 0.30~nm and 0.74~nm, respectively. Narrow-band filters of this FWHM provide, 
as  already mentioned in Sect.~2, a good compromise between parasitic light, 
transmission, and signal-to-noise. The overall shape and the width of the interference 
filters at normalized intensities $I / I_0 = 0.5$, 0.1, and 0.01 match t
he specifications provided by the manufacturer (Barr Associates). This type of narrow-band 
interference filters is commonly used in combination with etalons. This is the reason 
why we did not use a specific central wavelength $\lambda_0$ in Fig.~\ref{FIG09}.

A key parameter to characterize an etalon is the finesse ${\cal F}$, which
is basically the free spectral range (FSR) divided by the FWHM of the
transmission profile. In the simplest form, the finesse ${\cal F}$ can be
related to the reflectivity $R$.

\begin{figure}[t]
\centerline{\includegraphics[width=\columnwidth]{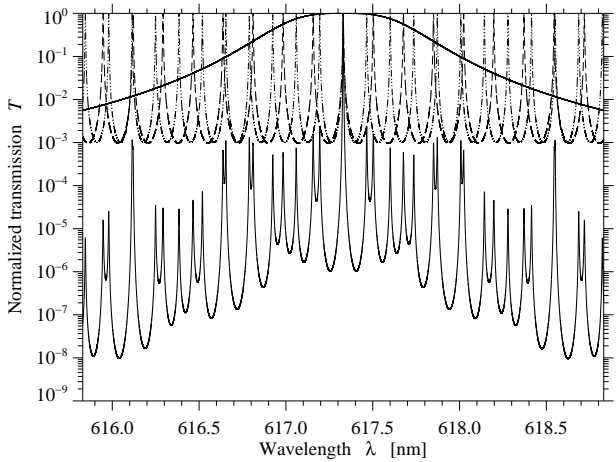}}
\caption{Transmission profile of the GFPI at Fe\,\textsc{i} $\lambda
    617.33 \pm 1.50$~nm. The curves correspond to a narrow-band 
interference filter ($\rm{FWHM} = 0.74$~nm, thick solid line), etalon 1 ($R = 0.95$, 
$d = 1.4$~mm, $\rm{FWHM} = 2.1$~pm, thin dashed line), etalon 2
($R = 0.95$, $d = 1.1$~mm, $\rm{FWHM} = 2.7$~pm, thin dash-dotted line), and 
the total transmission curve ($\rm{FWHM} = 1.5$~pm). The parasitic-light 
fraction of the combined filters is about 4.6\%.}
\label{FIG10}
\end{figure}

Higher order effects such as warping and shrinkage of the coatings are not taken
into account in the following calculations, where we assume a constant $p =
140$. The transmission profiles of the individual etalons and the order-sorting
interference filter are shown in Fig.~\ref{FIG10} for the Fe\,\textsc{i}
$\lambda 617.3$~nm line. The final GFPI transmission profile (thin solid curve)
was already multiplied by the transmission curve of the pre-filter (thick solid
curve). Out-off band transmission exceeds $I / I_0 = 10^{-3}$ only in a few
locations. Typically, secondary transmission peaks reach values of $I / I_0$ in
the order of $10^{-3}$--$10^{-5}$. The parasitic-light fraction is defined as the 
ratio of the intensity transmitted out-off band to that of the central transmission peak 
and has been calculated according to the wavelength range of the pre-filters shown in Fig.~\ref{FIG09}, 
i.e., $\pm 3$~nm and $\pm 4$~nm, respectively. In the case presented in Fig.~\ref{FIG10}, 
it amounts to about 4.6\% because of the relatively wide Fe\,\textsc{i} $\lambda 617.3$~nm 
pre-filter ($\mathrm{FWHM} = 0.75$~nm) with high transmission above 80\%.

\begin{figure}[t]
\centerline{\includegraphics[width=\columnwidth]{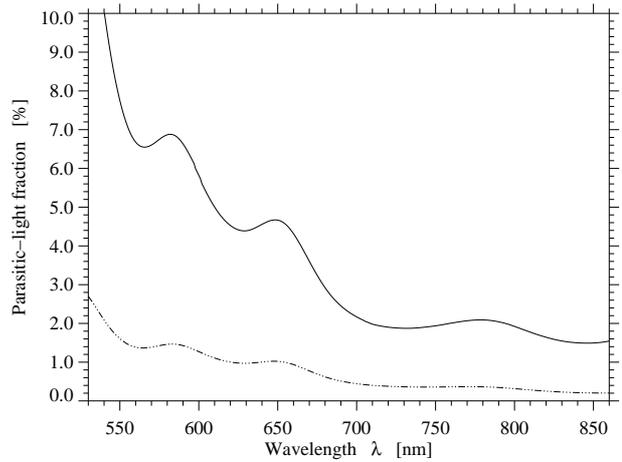}}
\caption{GFPI parasitic-light fraction for order-sorting interference filters
    with a FWHM of 0.75~nm (\textit{solid}\tsp) and 0.30~nm
    (\textit{dash-dotted}).}
\label{FIG11}
\end{figure}

The parasitic-light fraction can be minimized during the design of 
a dual- or multi-etalon system. However, when the preparations of the 
GFPI for the GREGOR telescope started in 2005, the first narrow-band etalon with a 
plate spacing of $d=1.1$\,mm was purchased in addition to an already 
existing broad-band etalon with $d=0.125$\,mm. The second narrow-band 
etalon with $d=1.4$\,mm was then purchased later on as the best 
choice in combination with the etalon with $d=1.1$\,mm, although in 
principal these plate spacings are not optimal. In Fig.~\ref{FIG11}, we present  
the wavelength-dependent parasitic-light fraction for the two 
interference filters in Fig.~\ref{FIG09} and for the characteristic 
parameters of the dual-etalon system. The two curves represent upper and lower
bounds for all narrow-band GFPI interference filters. The wavy appearance of the
curves has its origin in the coating curves (reflectivity $R_{1,2} (\lambda)$
and absorption $A_{1,2} (\lambda)$) of the etalons. In the case of the 0.75~nm
filter, one has to consider the trade-off between high-transmission on the one
side and high off-band contributions to the spectral line profile. Below 610~nm
the parasitic-light fraction exceeds 5\%, which might
no longer be acceptable for spectral inversions. Nevertheless, interference filters 
with a FWHM of 0.3~nm are well-suited to bring the parasitic-light fraction down to 
much lower values, (e.g., Fig.~\ref{FIG11} and \citet[][]{Puschmannetal2012b}, 
their Fig. 9, left panel). Nowadays, it is also possible to achieve a transmission above 60\% for
the narrower pre-filters but at additional cost. The choice of a suitable
pre-filter is ultimately driven by the science case of GFPI observations.

The spectral resolving power ${\cal R}$ of an FPI can be expressed as
\begin{equation}
{\cal R} = \frac{\lambda}{\Delta\lambda} = 0.97 m {\cal F},
\end{equation}
where $m$ is the order of interference \citep{Born1998}. If the angle of
incidence $\theta$ is almost normal to the etalon plates, the spectral
resolution becomes
\begin{equation}
{\cal R} = \frac{\lambda}{\Delta\lambda} \approx \frac{2 {\cal F} nd}{\lambda}
    \quad \mathrm{with} \quad \cos\theta \approx 1.
\end{equation}
Using Eq.~\ref{FIN}, we can relate the FSR to the bandpass $\Delta\lambda$ and
Finesse ${\cal F}$ by
\begin{equation}
\mathrm{FSR} = \frac{\lambda^2}{2nd} = \Delta\lambda {\cal F}.
\end{equation}

\begin{figure}[t]
\centerline{\includegraphics[width=\columnwidth]{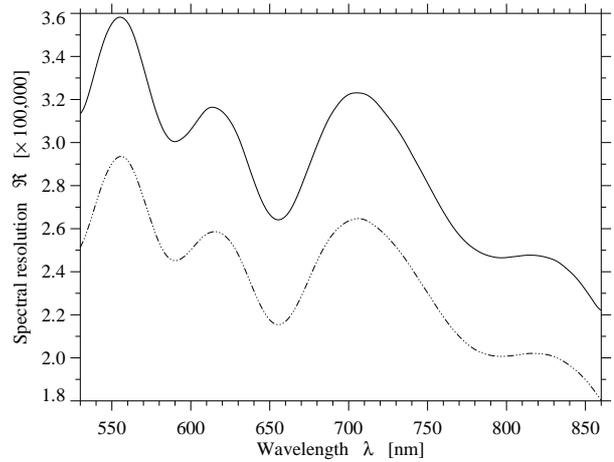}}
\caption{Spectral resolution ${\cal R}$ as a function of wavelength $\lambda$
     based on the FWHM \mbox{(\textit{solid}\tsp)} and equivalent width
     \mbox{(\textit{dash-dotted}\tsp)} of the dual-etalon system.}
\label{FIG12}
\end{figure}

This approach can be used for individual etalons. However, to compute the
wavelength-dependent spectral resolution ${\cal R} (\lambda)$, we first
multiplied the individual transmission curves, before computing the FWHM and the
equivalent width, i.e., the width of a box-type filter, which transmits the same
number of photons. Both curves are presented in Fig.~\ref{FIG12}. The extended
wings of the Airy function result in a larger equivalent width than the FWHM and
thus to a lower spectral resolution ${\cal R}$. In both cases, however, we
assume no higher-order effects, which can affect the finesse ${\cal F}$. The
highest spectral resolution can be achieved at shorter wavelengths. Again, the
shape of the curves has its origin in the coating curves of the etalons. If the
finesse changes across the pupil, then the spectral resolution can significantly
decrease for almost fully illuminated etalon plates.

During science verification, the dispersion and spectral resolution of the GFPI was determined \citep[][their Tab.~3]{Puschmannetal2012b}. 
In agreement with earlier measurements, the dispersion follows a linear relation 
\begin{equation}
\delta\lambda_{\rm min}=C_{0}+C_{1}\lambda\,,
\end{equation}
with the wavelength given in meters and the coefficients $C_{0}=0.93\times10^{-14}$\,m and $C_{1}=4.63\times10^{-7}$. The minimum step width 
at 530 and 860\,nm corresponds thus to 0.25 and 0.40 pm and yields a maximum scan range of 1.02 and 1.63 nm at these edge wavelengths, 
respectively, given the 12-bit resolution of the CS100 ICOS etalon controllers. 
With respect to the spectral resolution, we found a significant deviation between the theoretical (${\cal R}=250,000$) and 
effective spectral resolution (${\cal R}=100,000$) using the method described by \citet{Allendeprietotal2004} and \citet{Cabrerasolanaetal2007} that estimates
the latter by a comparison to a reference profile. At the VTT, \citet{PuschmannBeck2011} still
obtained a spectral resolution of ${\cal R} \approx$ 160,000 using the same method. However, 
the pupil image was smaller at the VTT ($\approx 40$~mm) than at the GREGOR telescope ($\sim 60$~mm). A degradation of the finesse between
the central and peripheral etalon regions of about 1.3 was found already for a single etalon for radii of 15~mm and 50~mm by \citet{DenkerTritschler2005}.

One of the advantages of a FPI in a collimated mounting is that each point
in the image plane has the same spectral resolution everywhere
\citep{Cavallini2006}, which is also higher as compared to similar FPIs in
telecentric mounting. In other words, each point in the images essentially
receives light from the entire pupil, i.e., any local defect of the plates will
diminish the finesse, thus leading to a lower spectral resolution. One approach
to mitigate this problem is to widen the beam diameter of the laser (currently
15~mm), which is used to maintain the parallelism of the plates. Sampling a
larger area of the etalon plates would result in a better compromise between the
contributions by the inner and peripheral parts of the etalon plates to the
overall finesse. The application of the above mentioned method to spectra 
from other spectrometers based on air-spaced etalons would be extremely 
interesting for comparison.

\begin{figure}[]
\centering
\includegraphics[width=\columnwidth]{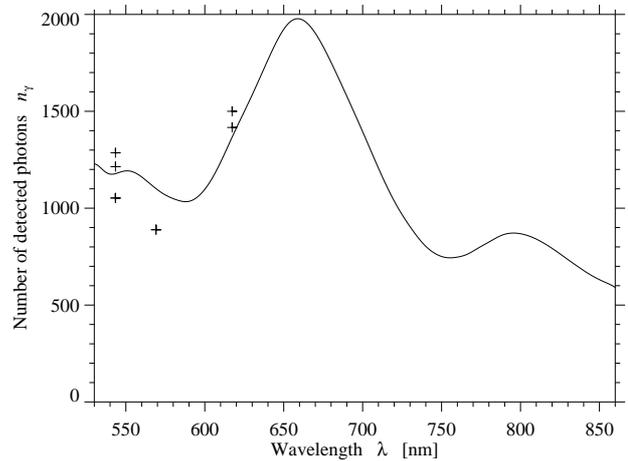}
\caption{Number of detected photons $n_\gamma$ as a function of wavelength
    $\lambda$ for an exposure time of $\Delta t = 10$~ms considering the
    transmission profile of the dual-etalon system, the quantum efficiency
    of the Sensicam QE camera, and all relevant characteristic parameters of the
    GFPI (without the transmission of the order-sorting pre-filter). The plus signs
    refer to actual measurements of the continuum intensity.}
\label{FIG13}
\end{figure}

Reducing the diameter of the pupil image is not a solution to circumvent
the above problem, because this would increase the blueshift
$\delta\lambda_\theta$ contained in the filtergrams, where the wavelength
information of each pixel is field-dependent \citep[see][]{Cavallini2006}
according to
\begin{equation}
\delta\lambda_\theta = - \frac{\lambda\theta^2}{2n},
\end{equation}
where $\theta$ is the angle of incidence and $n$ the index of refraction. If the
angle of incidence remains small, the Helmholtz-Lagrange invariant yields
\begin{equation}
\beta = \frac{D_\mathrm{FPI}}{f_\# D_\mathrm{tel}},
\end{equation}
with the field angle $\beta$, the $f$-number $f_\#$ of the optical system, and
the diameters of etalons $D_\mathrm{FPI}$ and telescope aperture
$D_\mathrm{tel}$. Combining these equations provides the expression
\begin{equation}
\delta\lambda_\theta = - \frac{\lambda}{8n} \left( \frac{\beta
D_\mathrm{tel}}{D_\mathrm{FPI}} \right)^2,
\end{equation}
which can be used to calculate the  blueshift $\delta\lambda_\theta =
3.7$--6.1~pm for the wavelength range 530--860~nm and a field angle $\beta =
60\arcsec$. Theoretical values and actual measurements of the maximal blueshift 
across the FOV at wavelengths between 543.4~nm and 630.2~nm are in very good agreement 
\citep[see][their Tab.~4]{Puschmannetal2012b}.

In planning an observing sequence, the number of detected photons $n_\gamma$
should be known as a function of wavelength $\lambda$. Extracting the bandpass
from the curves shown in Fig.~\ref{FIG12}, we can adjust this curve to match
GFPI observations as demonstrated in Fig.~\ref{FIG13}. The overall efficiency of
the telescope and other optical components was assumed to be constant at about
4.8\% to reach a good agreement between observations and theory. Note that the
observations might be affected by the sky transparency, the zenith distance of
the Sun, and to a lesser extent by the scene observed on the solar surface. We
have excluded the order-sorting interference filter from this comparison, which
would further reduce the number of detected photons $n_\gamma$, because its
characteristics strongly varies with the spectral line chosen for the
observations. Typically, a level of $n_\gamma \approx 6000$ photons should be
measured in the continuum of a spectral line, i.e., the exposure time $\Delta t$
and/or the binning have to be adjusted accordingly. In the last computation we
took into account a factor of two for the conversion of detected photons in
measured electrons in the analog-digital conversion. Thus, the continuum
intensity of the quiet Sun would reach about three quarters of the full-well
capacity of the Sensicam~QE detector. Measured values obtained by
\citet{Puschmannetal2012b} are in good agreement (see the corresponding plus signs in Fig.~\ref{FIG13}).

\section {Polarimetry}

\begin{figure}[]
\centering
\includegraphics[width=\columnwidth]{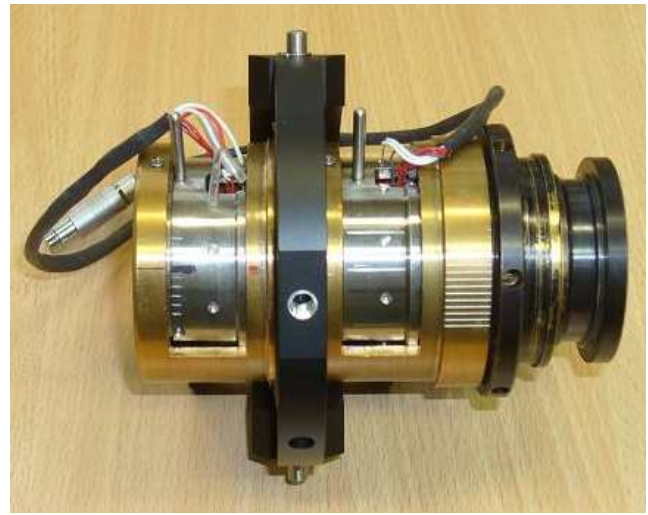}
\caption{Image of the Stokes vector polarimeter.}
\label{FIG14}
\end{figure}

A  polarimeter can be inserted just in front of the detector in the \textsf{NBC}
for using the GFPI in the dual-beam vector spectropolarimetric mode.
Both orthogonal beams are imaged on the same chip, 
thus reducing the available FOV to  $25\arcsec \times 38\arcsec$. 
The polarimeter by courtesy of the IAC is described in detail in \citet{BelloGonzalezKneer2008}. 
It consists of two ferro-electric liquid crystals (FLC) and two polarizing
beamsplitters, i.e., calcites. The latter split the light into ordinary and
extraordinary beam, which are separated by 4.2~mm. A half-wave plate between the
two calcites exchanges both beams. The first FLC acts roughly as a half-wave
plate and the second as a quarter-wave plate. FLCs are switched between two
states, which allows one to retrieve the full Stokes vector from four
independent measurements. An image of the polarimeter is shown in
Fig.~\ref{FIG14}. The FLCs in the silver-colored mounts are located inside the
main brass-colored housing and can be rotated against each other. The various
cables connect to the power supply and conduct the switch signals to the FLCs.
The calcites are mounted near the black-painted exit of the polarimeter to the
right. The central black ring connects the polarimeter to the mechanical holder
on the optical table.

\begin{figure}[t]
\centering
\includegraphics[width=\columnwidth]{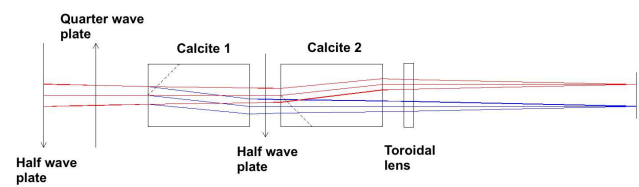}
\caption{ZEMAX ray-tracing including polarimeter and a toroidal lens with with a
    curvature radius of 710~mm as corrector for the astigmatism introduced by
    the polarimeter. From left to right: dual-beam polarimeter with FLC1 ($\lambda/2$),
    FLC2 ($\lambda/4$), calcite 1, half-wave plate, calcite 2, and at its exit
    the toroidal lens. Red and blue rays show ordinary and extraordinary beams, respectively.}
\label{FIG15}
\end{figure}

\begin{figure*}
\begin{center}
\includegraphics[width=\columnwidth]{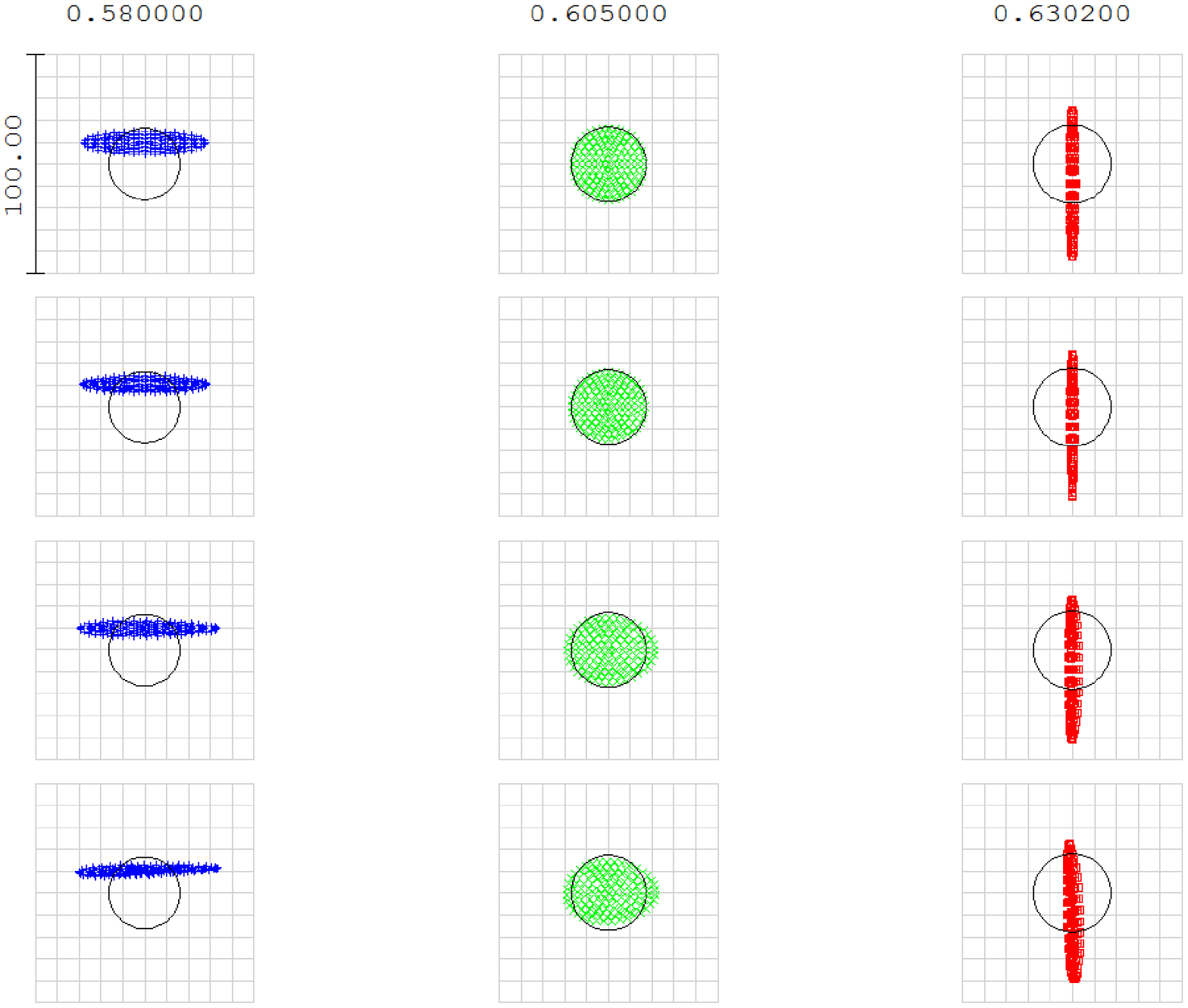}
\includegraphics[width=\columnwidth]{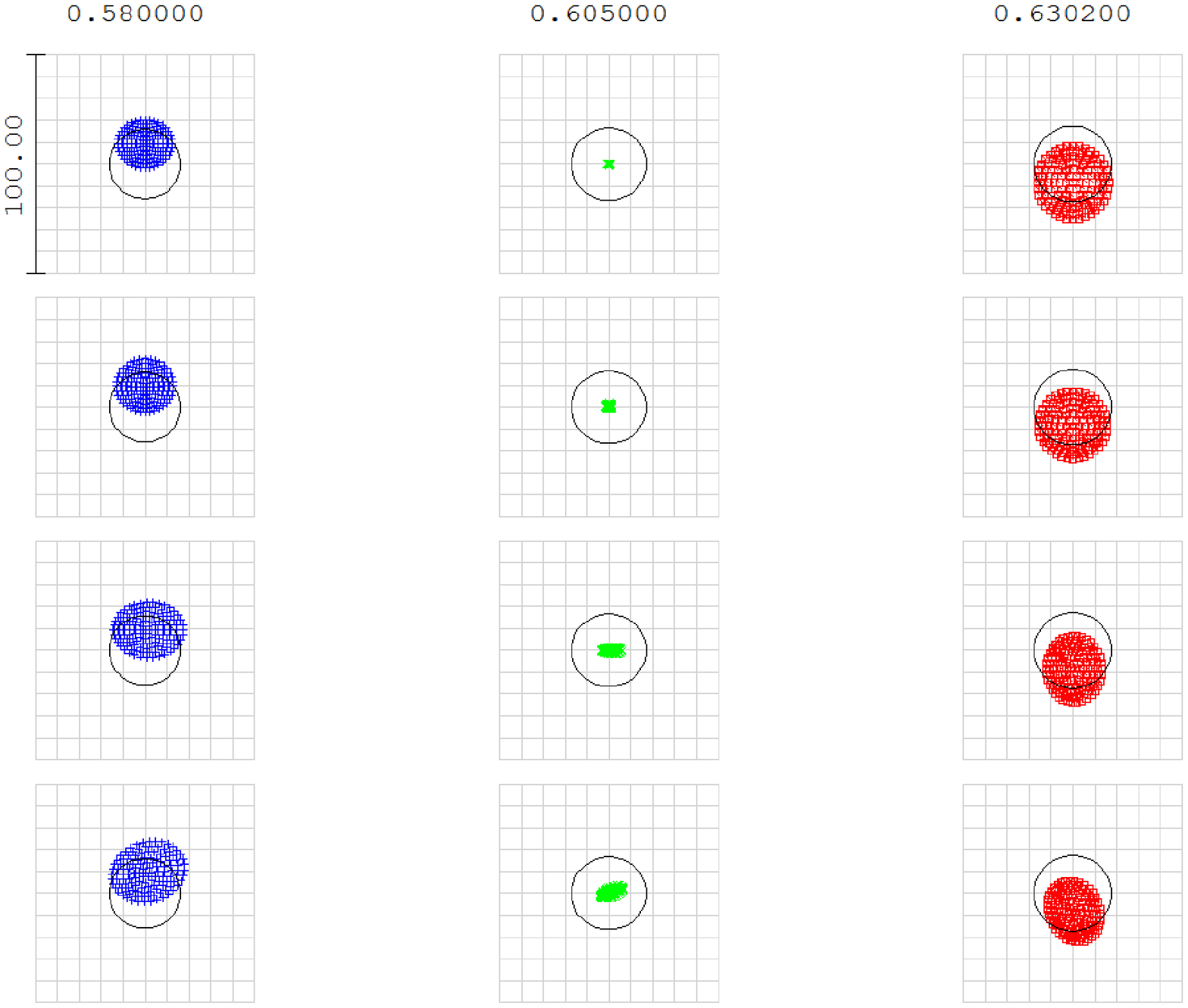}
\end{center}
\caption{Matrix spot diagrams (wavelength range 580--630~nm with a focus on
    605~nm) for the NBC including the Stokes vector polarimeter. Left panel:
    without corrector. Right panel: the same including a toroidal lens with a
    curvature radius of 710~mm as depicted in Fig.~\ref{FIG15} as corrector.
    From top to bottom: central optical axis, and at field positions of 
    $x=0$~mm, $y=3$~mm; $x=5.3$~mm, $=0$~mm;  $x=5.3$~mm, $y=3$~mm.}
\label{FIG16}
\end{figure*}

The projected lengths of the  extraordinary beam in the $xy$- and $xz$-planes
are different when passing the first calcite, which introduces astigmatism. The
ordinary beam of the first calcite is then affected by the second calcite,
producing once more the same amount of astigmatism with the same orientation.
The astigmatism can be mostly compensated by a toroidal lens with a curvature
radius of 710~mm (see Figs.~\ref{FIG15} and \ref{FIG16}). The astigmatism caused
by double refraction is the dominating aberration followed by spherical
aberration and exceeds the Airy disk, which can be easily seen in the spot
diagrams for the peripheral wavelengths (left panel of Fig.~\ref{FIG16}). The
compensation for astigmatism and spherical aberration with a weak cylindrical
lens (one-dimensional toroid) faced with the curvature towards the calcites
compensates both aberrations and gives a diffraction-limited imaging for the
central wavelength and just diffraction-limited imaging for the peripheral
wavelengths (right panel of Fig.~\ref{FIG16}). The strength of the aberrations
is nearly independent of the field position.

The retardance  $\delta_{1,2}$ and cone angle $\alpha_{1,2}$ of the FLCs, and
the polarimetric efficiency have been measured in the laboratory at three
different wavelengths, i.e., 557~nm, 620~nm, and 656~nm (Table~\ref{Tab4}). An
optimum polarimetric efficiency resulted when the two FLCs were rotated by
$20^\circ$ relative to each other. The efficiencies of about 50\% are quite
good for 557~nm and 630~nm, but they are not satisfactory at 656~nm with
about 20\% efficiency only.

\begin{table}
\caption{Retardance $\delta_{1,2}$ and cone angle $\alpha_{1,2}$ for FLC~1 and
    FLC~2, and polarimetric efficiencies $\epsilon$.}
\label{Tab4}

\begin{tabular}{lccccc}
\hline\hline
FLC~1\rule[-1.5mm]{0mm}{5mm}  & $\delta_1$ & $\delta_2$ & $\alpha_1$ &
    $\alpha_2$    & $\Delta\alpha$ \\
\hline
557~nm\rule{0mm}{3.5mm}       & 144.0 & 139.5 & 51.3 & 11.8 & 39.5 \\
620~nm                        & 162.8 & 155.1 & 50.1 & \phn 9.9 & 40.2 \\
656~nm\rule[-1.5mm]{0mm}{3mm} & 163.3 & 160.8 & 50.5 & 10.0 & 40.5 \\
\hline
&&&&&\\
\hline\hline
FLC~2\rule[-1.5mm]{0mm}{5mm} & $\delta_1$ & $\delta_2$ & $\alpha_1$ & $\alpha_2$
& $\Delta\alpha$ \\
\hline \\
557~nm\rule{0mm}{3.5mm}  & 100.7 &  96.5 & 85.3 & 41.5 & 43.8 \\
620~nm  & \phn 83.0 &  80.4 & 86.5 & 40.5 & 46.0 \\
656~nm\rule[-1.5mm]{0mm}{3mm}  & \phn 78.4 &  80.4 & 58.4 & 17.6 & 40.8 \\
\hline
&&&&&\\
\hline\hline
$\epsilon$\rule[-1.5mm]{0mm}{5mm} & $\epsilon(I)$ & $\epsilon(Q)$ &
$\epsilon(U)$ & $\epsilon(V)$ & $\epsilon_\mathrm{total}$ \\
\hline \\
557~nm\rule{0mm}{3.5mm}  & 0.978  &   0.565 &  0.519  &  0.518 &  0.926 \\
620~nm  & 0.971  &   0.585 &  0.481  &  0.620 &  0.979 \\
656~nm\rule[-1.5mm]{0mm}{3mm}  & 0.430  &   0.191 &  0.200  &  0.366 &  0.459 \\
\hline
\end{tabular}
\end{table}

For accurate polarimetric measurements, a calibration of the complete optical
train up to the polarimeter must be performed
\citep{Collados2003,Becketal2005a,Becketal2005b}. For this purpose, the GREGOR
Polarimetric calibration Unit \citep[GPU,][]{Hofmannetal2009, Hofmannetal2012}
can be inserted close to the secondary focus F2 of the telescope. The GPU
polarizes the incident light in a defined and known way. However, the internal
rotations of the alt-azimuthally mounted telescope change the instrumental
polarization quite rapidly, therefore a time-dependent polarization model of
the GREGOR solar telescope was developed \citep{Balthasaretal2011,
Hofmannetal2012}. Together with regular calibration measurements during the
observing campaigns using the GPU, this model will serve to calibrate the
spectropolarimetric data taken with both the GFPI and GRIS.

However, the GPU can currently only be used for polarimetric calibrations
in the near-infrared wavelength range because of thermal issues with the
retarders at visible wavelengths. An additional protection with a partly
transparent mirror in front of the GPU is planned to reduce the heat load. A
second calibration unit \citep{Colladosetal2012} is foreseen for both the
visible and near-infrared spectral range. This unit will be inserted behind the
telescope exit in the observing room to determine the instrumental polarization
of the static optical path of AO system and the post-focus instruments. Until
the full availability of one or both calibration units in the next year, the
observations with the GFPI are restricted to the spectroscopic mode. The
implementation of an automated calibration procedure for the GFPI is expected to
be finished along with the upgrade of the calibration unit(s). An extension 
of the usable wavelength range of the polarimeter to 530--860~nm by 
the addition of one variable FLC retarder \citep[see][]{Tomczyketal2010} is under 
investigation. However, it is a major task to eventually enlarge the usable FOV in 
the spectropolarimetric mode. The replacement of the Sensicam cameras by modern, large-format 
sCMOS cameras as described in \citet{Puschmannetal2012b} would lead to both higher frame 
rates and a larger FOV. However, the larger FOV also requires a re-design or replacement 
of the polarimeter itself, because at present the FOV is also limited by the size of the calcites.

\section{Science with imaging spectropolarimeters}

Imaging spectropolarimeters have been used in solar physics for more than two
decades and produced a wealth of scientific results. Looking more closely at
previous scientific studies with imaging spectropolarimeters -- such as the
G\"ottingen Fabry-P\'erot Interferometer \citep{Bendlinetal1992} and the
Telecentric Etalon SOlar Spectrometer \citep[TESOS,][]{Tritschleretal2004a} at the
VTT, the CRISP instrument \citep{Scharmer2008} at the 1-meter Swedish Solar
Telescope, the Interferometric BIdimensional Spectrometer
\citep[IBIS,][]{Cavallini2006, ReardonCavallini2008} at the Dunn Solar Telescope, and the Imaging
Magnetograph eXperiment \citep[IMaX,][]{MartinezPillet2011} for the Sunrise
Balloon-Borne Solar Observatory -- reveals the strengths of these instruments to
easily adapt to science cases ranging from the quiet Sun to the dynamics of
sunspots while covering photospheric as well as chromospheric lines.

\begin{figure*}[t]
\begin{center}
\includegraphics[width=\textwidth]{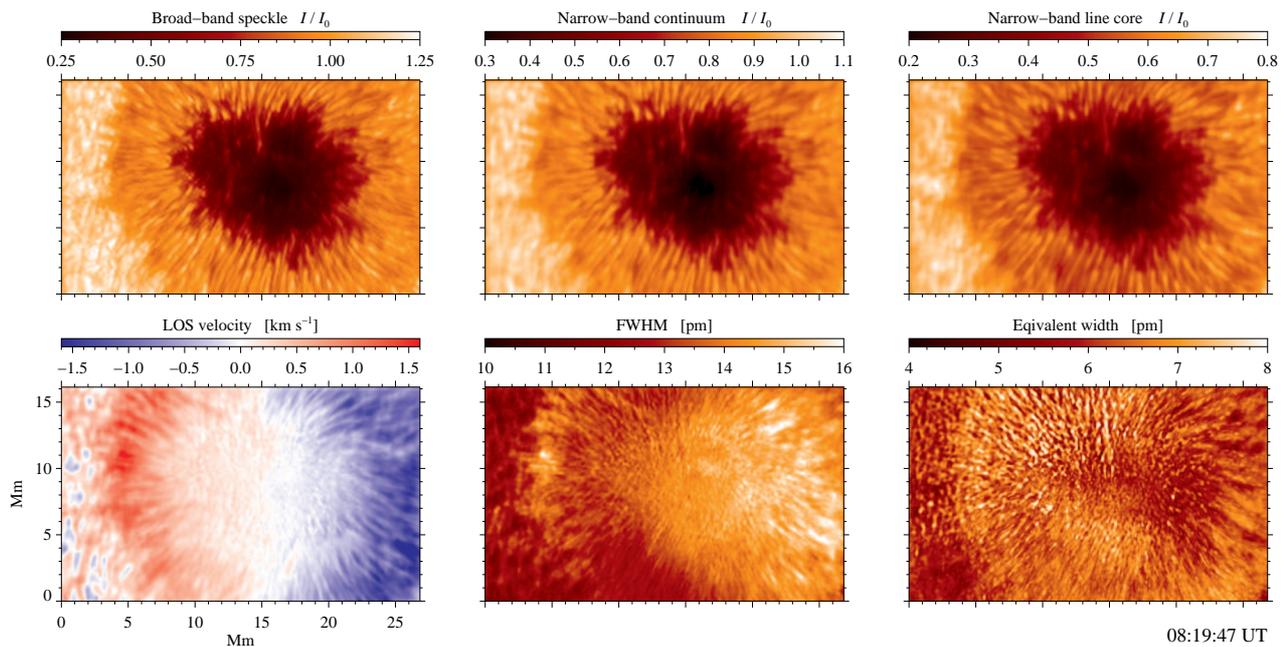}
\end{center}
\caption{Example snapshot of a 22-min time series of NOAA AR 11530.
         Upper panels: speckle-reconstructed broad-band image  at 620\,nm (left), speckle-deconvolved continuum (middle) and line-core (right) narrow-band images. 
         Bottom panels: line-core velocity (left), full-width at half-maximum (middle), and equivalent width (right).}
\label{FIG17}
\end{figure*}

The polarimetric accuracy of imaging spectropolarimeters is a few $10^{-3}$ in
terms of the continuum intensity, which is sufficient to follow the coalescence,
fragmentation, and cancellation of small-scale magnetic features.
\citet{Viticchie2009} studied their association with G-band bright points and
used the SIR (Stokes inversions based on response function) approach
\citep{RuizCobo1992} to determine that their temporal evolution is related to
the filling factor rather than to the magnetic field strength. Observations with
high polarimetric sensitivity revealed that magnetic fields occupy approximately
45\% of the area in internetwork patches, predominantly in intergranular spaces
\citep{DominguezCerdena2003}. Scrutinizing the intricate fine structure of
granules, \citet{Steiner2010} found evidence for horizontally oriented vortex
tubes, i.e., striations with leading bright rim and a trailing dark edge, which
migrate in unison from the periphery to the granule's interior.

The infrared Ca\,\textsc{ii} $\lambda 854.2$~nm line is an excellent diagnostics
tool \citep{Cauzzi2008}, because photospheric conditions can be examined in its
far wings, whereas the line core is exclusively of chromospheric origin.
High-cadence observations are a prerequisite to measure oscillations in the
photosphere, e.g., \citet{Vecchio2007} identified network magnetic elements,
which guide low-frequency oscillations from the photosphere into the
chromosphere. Type~II spicules have also received much attention, because of
their role in providing hot material to the chromosphere \citep{Langangen2008}
and thus contribute to the energy balance of the lower solar atmosphere.
High-speed jets were observed in the chromospheric H$\alpha$ $\lambda 656.3$~nm
($\approx 50$~km~s$^{-1}$) and infrared Ca\,\textsc{ii} $\lambda 854.2$~nm
($\approx 20$~km~s$^{-1}$) lines by \citet{RouppevanderVoort2009}, who consider
them as on-disk counterparts of type~II spicule.

Imaging spectropolarimetry of sunspots has been mainly concerned with the fine
structure of the penumbra. A comprehensive study of a regular sunspot with a
light-bridge and umbral dots was presented by \citet{Tritschleretal2004b}, who used
maps of the line-of-sight velocity, the line width, the equivalent width, and
the line depression for a comprehensive description of sunspot dynamics. The
elusive process of penumbra formation has been followed in detail in a four-hour
time-series presented in \citet{Schlichenmaier2010}. The ``uncombed'' nature of
penumbral filaments, their association with the Evershed flow, and their ``sea
serpent''-like extension beyond the sunspot's boundaries have been examined by
\citet{BelloGonzalez2005}. More recently, \citet{Scharmer2011} detected
ubiquitous downflows (up to 1.0~km~s$^{-1}$) in the inner penumbra and a strong
Evershed flow (3.0--3.5~km~s$^{-1}$) in the outer penumbra, pointing one more
time to the complex and very dynamic nature of sunspots.

The investigations described above only present a small cross-section of
potential science cases for imaging spectropolarimetry. Every time when
high-cadence and high-spatial resolution observations are required to unravel
the physics of dynamic solar features, the GFPI and its cousins are the
instruments of choice.

\section{First observations with adaptive optics}

Figure~\ref{FIG17} shows the results of one of the first two-dimensional 
spectral scans obtained at the GREGOR telescope with the GFPI and real-time correction 
by GAOS on 31 July 2012. A 22-min time series of the active region NOAA 11530 located 
at S24 W18 (heliocentric angle $\theta=30^{\circ}$) was observed between 8:04 and 8:26 UT. The angle between 
solar south and the center of symmetry was 33.8$^{\circ}$. Therefore, the direction from center 
to limb corresponds approximately to the $x$-axis. 
We scanned the Fe\,\textsc{i} $\lambda 617.3$~nm line with 86 steps and a step size of $\Delta\lambda= 1.17$~pm, 
taking eight narrow-band images ($\rm{FWHM}=2.1$~pm) at each wavelength position. Broad-band images ($\rm{FWHM}=10$\,nm) 
at 620~nm were simultaneously recorded. The exposure time was 20~ms and each line scan took 64~s. We used a $2\times2$ 
on-chip binning corresponding to a spatial sampling of 0\farcs072, which resulted in a total FOV of $\sim 36{\,\rm Mm} \times 27{\,\rm Mm}$.
The data reduction included, apart from routine tasks such as flatfielding, a speckle reconstruction 
of the broad-band data \citep{PuschmannSailer2006, deBoer1993} 
and a subsequent deconvolution of the narrow-band data with a modified version 
of the code by \citet{Janssen2003} that is based on the 
method of \citet{KellervdL1992}. The narrow-band data were corrected for the blueshift 
across the FOV after the deconvolution. Subsequently, the image rotation caused by 
the alt-azimuthal mount of the GREGOR telescope ($\sim$~6$^{\circ}$ in 20~min) was compensated and the data were spatially
aligned. Line parameters calculated from the spectra are shown in Fig.~\ref{FIG17}, together with the speckle-reconstructed broad-band image. 
The final FOV is reduced to $\rm{27~Mm}\times \rm{16~Mm}$ because of the data reduction described 
above and an additional apodisation using  a sub-sonic filter with a cutoff-velocity of 5~km~s$^{-1}$.

The contrast of the speckle-deconvolved spectral continuum and line-core narrow-band images is somewhat 
below that of the speckle-reconstructed  broad-band image, but individual umbral dots are still easy to identify. 
The line-of-sight velocity map clearly reveals the filamentary structure of the penumbra, with strong outflows related 
to intraspines and velocities close to zero in the spines.

As far as the spatial resolution is concerned, a minimization of the straylight problem currently present at the GREGOR telescope 
will improve significantly the performance of GAOS in the near future. At the same time, we try to improve the performance of both speckle 
and blind deconvolution methods by adjusting them to the specific conditions at GREGOR. Thus, we expect that very soon deconvolved 
spectra near the diffraction limit will be regularly available.

\section{Conclusions}

The GREGOR Fabry-P\'erot Interferometer is designed for imaging spectroscopy
(spectropolarimetry) in the range of 530--860~nm (580--660~nm) with a
diffraction-limited performance at a spatial sampling of 0\farcs036 and a
theoretical spectral resolution of ${\cal R} =$ 250,000.  Options 
for the extension of the wavelength range for polarimetry to 530--860~nm are under investigation.
The technical performance of the instrument was verified by direct measurements with the GFPI 
at the GREGOR telescope during science verification \citep{Puschmannetal2012b}. 
Most characteristic parameters of the instrument complied with the theoretical expectations. 
However, a strong deviation between theoretical and measured spectral resolution was found,
which will require some investigation before the next observing campaign. The above mentioned 
measurements also revealed the need for improving the total throughput for spectropolarimetric 
observations by either using high-transmission (80\%) pre-filters or a $2 \times 2$ on-chip 
binning until new cameras will be available.

The first 2D-spectra obtained with AO support by the end of July 2012 show a very good performance of the GFPI. 
The instrument is ready, awaiting its users with automatic observing sequences that provide an easy access to 
scientific observations. An almost completed data reduction pipeline, which also includes image restoration 
techniques, will further provide an easy handling of the GFPI high-spatial resolution data.


\acknowledgements The 1.5-meter GREGOR solar telescope was build by a German 
consortium under the leadership of the Kiepenheuer-Institut f\"ur Sonnenphysik 
in Freiburg with the Leibniz-Institut f\"ur Astrophysik Potsdam, 
the Institut f\"ur Astrophysik G\"ottingen, and the Max-Planck-Institut f\"ur 
Sonnensystemforschung in Katlenburg-Lindau as partners, and with contributions 
by the Instituto de Astrof\'\i sica de Canarias and the Astronomical Institute 
of the Academy of Sciences of the Czech Republic. CD was supported by
grant DE~787/3-1 of the Deutsche Forschungsgemeinschaft (DFG). We would like to
thank Robert Geissler for his competent and extended help during the science
verification.


\begin{thebibliography}{99}

\bibitem[\protect\citeauthoryear{Allende Prieto et al.}{2004}]{Allendeprietotal2004}
Allende Prieto, C., Asplund, M., \& Fabiani Bendicho, P.: 2004, A\&A 423, 1109
\bibitem[\protect\citeauthoryear{Balthasar et al.}{2009}]{Balthasaretal2009}
    Balthasar, H., Bello Gonz{\'a}lez, N., Collados, M., et al.: 2009, in: K.G.\
    Strassmeier, A.G.\ Kosovichev, J.E.\ Beckman (eds.), \textit{Cosmic Magnetic
    Fields: From Planets, to Stars and Galaxies}, IAU Symp.\ 259, 665
\bibitem[\protect\citeauthoryear{Balthasar et al.}{2011}]{Balthasaretal2011}
    Balthasar, H., Bello Gonz{\'a}lez, N., Collados, M., et al.: 2011, in: J.R.\
    Kuhn et al.\ (eds.), \textit{Solar Polarization Workshop 6}, ASPC 437, 351
\bibitem[\protect\citeauthoryear{Beck et al.}{2005a}]{Becketal2005a}
    Beck, C., Schlichenmaier, R., Collados, M., et al.: 2005a, A\&A 443, 1047 
\bibitem[\protect\citeauthoryear{Beck et al.}{2005b}]{Becketal2005b}
    Beck, C., Schmidt, W., Kentischer, T., \& Elmore, D.: 2005b, A\&A 437, 1159 
\bibitem[\protect\citeauthoryear{Berkefeld et al.}{2012}]{Berkefeldetal2012}
    Berkefeld, T., Schmidt, D., Soltau, D., et al.: 2012, AN 333, 863
\bibitem[\protect\citeauthoryear{Bendlin et al.}{1992}]{Bendlinetal1992}
    Bendlin, C., Volkmer, R., \& Kneer, F.: 1992, A\&A 257, 817
\bibitem[\protect\citeauthoryear{Bello Gonz{\'a}lez \&
    Kneer}{2008}]{BelloGonzalezKneer2008}
    Bello Gonz{\'a}lez, N. \& Kneer, F.: 2008, A\&A 480, 265
\bibitem[\protect\citeauthoryear{Bello Gonz\'alez et
    al.}{2005}]{BelloGonzalez2005}
    Bello Gonz\'alez, N., Okunev, O.V., Dom\'{\i}nguez Cerde\~na, I., et al.:
    2005, A\&A 434, 317
\bibitem[\protect\citeauthoryear{Born \& Wolf}{1998}]{Born1998}
    Born, M. \& Wolf, E.: 1998, \textit{Principles of Optics}, 6$^{\mathrm{th}}$
    ed., University Press, Cambridge
\bibitem[\protect\citeauthoryear{Cabrera Solana et al.}{2007}]{Cabrerasolanaetal2007}
Cabrera Solana, D., Bellot Rubio, L. R., Beck, C., et al.: 2007, A\&A 475, 1067
\bibitem[\protect\citeauthoryear{Cauzzi et al.}{2008}]{Cauzzi2008}
    Cauzzi, G., Reardon, K.P., Uitenbroek, H., et al.: 2008, A\&A 480, 515
\bibitem[\protect\citeauthoryear{Cavallini}{2006}]{Cavallini2006}
    Cavallini, F.: 2006, SoPh 236, 415  
\bibitem[\protect\citeauthoryear{Collados}{2003}]{Collados2003}
    Collados, M.: 2003, in: S.\ Fineschi (ed.), \textit{Polarimetry in
    Astronomy}, Proc.\ SPIE 4843, 55
\bibitem[\protect\citeauthoryear{Collados et al.}{2008}]{Colladosetal2008}
    Collados, M., Calcines, A., D{\'{\i}}az, J.J., et al.: 2008, in: I.S.\
    McLean, M.M.\ Casali (eds.), \textit{Ground-Based and Airborne
    Instrumentation for Astronomy II}, Proc.\ SPIE 7014, 5Z
\bibitem[\protect\citeauthoryear{Collados et al.}{2012}]{Colladosetal2012}
    Collados, M., L\'opez, R., P\'aez, E., et al.: 2012, AN 333, 872
\bibitem[\protect\citeauthoryear{de Boer}{1993}]{deBoer1993}
de Boer, C.R.: 1993, PhD Thesis, Georg-August Universit\"at G\"ottingen, Germany
\bibitem[\protect\citeauthoryear{Denker \& Tritschler}{2005}]{DenkerTritschler2005}	
    Denker, C. \& Tritschler, A.: 2005, PASP 117, 1435
\bibitem[\protect\citeauthoryear{Denker}{2010}]{Denker2010a}
    Denker, C.: 2010, AN 331, 648
\bibitem[\protect\citeauthoryear{Denker et al.}{2010}]{Denkerteal2010b}
    Denker, C., Balthasar, H., Hofmann, A., et al.: 2010, in: I.S.\ McLean,
    S.K.\ Ramsay, H.\ Takami (eds.), \textit{Ground-Based and Airborne
    Instrumentation for Astronomy III}, Proc.\ SPIE 7735, 6M
\bibitem[\protect\citeauthoryear{Dom\'{\i}nguez Cerde\~na et
    al.}{2003}]{DominguezCerdena2003}
    Dom\'{\i}nguez Cerde\~na, I., Kneer, F., \& S\'anchez Almeida, J.: 2003, ApJL
    582, 55
\bibitem[\protect\citeauthoryear{Halbgewachs et al.}{2012}]{Halbgewachs2012}
    Halbgewachs, C., Caligari, P., Glogowski, K., et al.: 2012, AN 333, 840
\bibitem[\protect\citeauthoryear{Hofmann et al.}{2009}]{Hofmannetal2009}
    Hofmann, A., Rendtel, J., Arlt, K.: 2009, Centr.\ Eur.\ Astrophys.\ Bull.\
    33, 317
\bibitem[\protect\citeauthoryear{Hofmann et al.}{2012}]{Hofmannetal2012}
    Hofmann, A., Arlt, K., Balthasar, H., et al.: 2012, AN 333, 854
\bibitem[\protect\citeauthoryear{Kneer \&
    Hirzberger}{2001}]{KneerHirzberger2001}
    Kneer, F. \& Hirzberger, H.: 2001, AN 322, 375
\bibitem[\protect\citeauthoryear{Janssen}{2003}]{Janssen2003}
Janssen, K.: 2003, PhD Thesis, Georg-August Universit\"at G\"ottingen, Germany
\bibitem[\protect\citeauthoryear{Keller \& von der L\"uhe}{1992}]{KellervdL1992}
Keller, C.U. \& von der L\"uhe, O.: 1992, A\&A 261, 321
\bibitem[\protect\citeauthoryear{Kneer et al.}{2003}]{Kneeretal2003}
    Kneer, F., Al, N., Hirzberger, J., Nicklas, H., \& Puschmann, K.G.: 2003, AN
    324, 302
\bibitem[\protect\citeauthoryear{Koschinsky et al.}{2001}]{Koschinskyetal2001}
    Koschinsky, M., Kneer, F., \& Hirzberger, J.: 2001, A\&A 365, 588
\bibitem[\protect\citeauthoryear{Langangen et al.}{2008}]{Langangen2008}
    Langangen, {\O}., de Pontieu, B., Carlsson, M., et al.: 2008, ApJL 679, 167
\bibitem[\protect\citeauthoryear{Mart\'{\i}nez Pillet et
    al.}{2011}]{MartinezPillet2011}
    Mart\'{\i}nez Pillet, V., del Toro Iniesta, J.C., \'Alvarez-Herrero, A., et
    al.: 2011, SoPh 268, 57
\bibitem[\protect\citeauthoryear{Puschmann et al.}{2006}]{Puschmannetal2006}
    Puschmann, K.G., Kneer, F., Seelemann, T., \& Wittmann, A.D.: 2006, A\&A 451,
    1151
\bibitem[\protect\citeauthoryear{Puschmann \& Sailer}{2006}]{PuschmannSailer2006}
    Puschmann, K.G.\& Sailer, M.: 2006, A\&A 454, 1151
\bibitem[\protect\citeauthoryear{Puschmann et al.}{2007}]{Puschmannetal2007}
    Puschmann, K.G., Kneer, F., Nicklas, H., \& Wittmann, A.D.: 2007, in: F.\
    Kneer, K.G.\ Puschmann, A.D.\ Wittmann (eds.), \textit{Modern Solar
    Facilities -- Advanced Solar Science}, p.~45
\bibitem[\protect\citeauthoryear{Puschmann et al.}{2012a}]{Puschmannetal2012a}
    Puschmann, K.G., Balthasar, H., Bauer, S.-M., et al.: 2012a, in:  T.R.\
    Rimmele (ed.), \textit{Magnetic Fields from the Photosphere to the Corona},
    ASPC 463, in press, \EDT{ArXiv e-prints, 1111.5509}
\bibitem[\protect\citeauthoryear{Puschmann et al.}{2012b}]{Puschmannetal2012b}
    Puschmann, K.G., Balthasar, H., Beck, C., et al.: 2012b, in: I.S.\ McLean,
    S.K.\ Ramsay, H.\ Takami (eds.), \textit{Ground-Based and Airborne
    Instrumentation for Astronomy IV}, Proc.\ SPIE 8446, in press, \EDT{ArXiv e-prints,
    1207.2084}
\bibitem[\protect\citeauthoryear{Puschmann \& Beck}{2011}]{PuschmannBeck2011}
    Puschmann, K.G. \& Beck, C.: 2011, A\&A 533, A21
\bibitem[\protect\citeauthoryear{Reardon \& Cavallini}{2008}]{ReardonCavallini2008}
    Reardon, K.P. \& Cavallini, F.: 2008, A\&A 481, 897
\bibitem[\protect\citeauthoryear{Rouppe van der Voort et
    al.}{2009}]{RouppevanderVoort2009}
    Rouppe van der Voort, L., Leenaarts, J., de Pontieu, B., et al.: 2009, ApJ
    705, 272
\bibitem[\protect\citeauthoryear{Ruiz Cobo \& del Toro
    Iniesta}{1992}]{RuizCobo1992}
    Ruiz Cobo, B. \& del Toro Iniesta, J.C.: 1992, ApJ 398, 375
\bibitem[\protect\citeauthoryear{Scharmer et al.}{2011}]{Scharmer2011}
    Scharmer, G.B., Henriques, V.M.J., Kiselman, D., \& de la Cruz Rodr\'{\i}guez,
    J.: 2011, Science 333, 316
\bibitem[\protect\citeauthoryear{Scharmer et al.}{2008}]{Scharmer2008}
    Scharmer, G.B., Narayan, G., Hillberg, T., et al.: 2008, ApJL 689, 69
\bibitem[\protect\citeauthoryear{Schlichenmaier et
    al.}{2010}]{Schlichenmaier2010}
    Schlichenmaier, R., Rezaei, R., Bello Gonz\'alez, N., \& Waldmann, T.A.: 2010,
    A\&A 512, L1
\bibitem[\protect\citeauthoryear{Schmidt et al.}{2012a}]{Schmidtetal2012a}
    Schmidt, W., von der L{\"u}he, O., Volkmer, R., et al.: 2012a, in: T.R.\
    Rimmele, (ed.) \textit{Magnetic Fields from the Photosphere to the Corona},
    ASPC 463, in press, \EDT{ArXiv e-prints, 1202.4289}
\bibitem[\protect\citeauthoryear{Schmidt et al.}{2012b}]{Schmidtetal2012b}
    Schmidt, W., von der L{\"u}he, O., Volkmer, R., et al.: 2012b, AN 333, 796
\bibitem[\protect\citeauthoryear{Steiner et al.}{2010}]{Steiner2010}
    Steiner, O., Franz, M., Bello Gonz\'alez, N., et al.: 2010, ApJL 723, 180
\bibitem[\protect\citeauthoryear{Tomczyk et al.}{2010}]{Tomczyketal2010}
Tomczyk, S., Casini, R., de Wijn, A.G., et al.: 2010, Ap.Opt. 49, 3580
\bibitem[\protect\citeauthoryear{Tritschler et al.}{2004a}]{Tritschleretal2004a}
    Tritschler, A., Bellot Rubio, L.R., \& Kentischer, T.J.: 2004a, AAS 204, 6902
\bibitem[\protect\citeauthoryear{Tritschler et al.}{2004b}]{Tritschleretal2004b}
    Tritschler, A., Schlichenmaier, R., Bellot Rubio, L.R., et al.: 2004b, A\&A
    415, 717
\bibitem[\protect\citeauthoryear{Vecchio et al.}{2007}]{Vecchio2007}
    Vecchio, A., Cauzzi, G., Reardon, K.P., et al.: 2007, A\&A  461, 1
\bibitem[\protect\citeauthoryear{Viticchi\'e et al.}{2009}]{Viticchie2009}
    Viticchi\'e, B., del Moro, D., Berrilli, F., et al.: 2009, ApJL 700, 145
\bibitem[\protect\citeauthoryear{Volkmer et al.}{1995}]{Volkmeretal1995}
    Volkmer, R., Kneer, F., \& Bendlin, C.: 1995, A\&A  304,  L1
\bibitem[\protect\citeauthoryear{Volkmer et al.}{2010}]{Volkmer2010}
    Volkmer, R., von der L{\"u}he, Denker, C., et al.: 2010, AN 331, 624
\bibitem[\protect\citeauthoryear{von der L\"uhe et al.}{2012}]{VDL2012}
    von der L\"uhe, O., Volkmer, R., Kentischer, T., et al.: 2012, AN 333, 894
\end{thebibliography}
\end{document}